\documentclass[article,nojss]{jss}\usepackage[]{graphicx}\usepackage[]{color}
\makeatletter
\def\maxwidth{ %
  \ifdim\Gin@nat@width>\linewidth
    \linewidth
  \else
    \Gin@nat@width
  \fi
}
\makeatother

\definecolor{fgcolor}{rgb}{0.345, 0.345, 0.345}
\newcommand{\hlnum}[1]{\textcolor[rgb]{0.686,0.059,0.569}{#1}}%
\newcommand{\hlstr}[1]{\textcolor[rgb]{0.192,0.494,0.8}{#1}}%
\newcommand{\hlopt}[1]{\textcolor[rgb]{0,0,0}{#1}}%
\newcommand{\hlstd}[1]{\textcolor[rgb]{0.345,0.345,0.345}{#1}}%
\newcommand{\hlkwa}[1]{\textcolor[rgb]{0.161,0.373,0.58}{\textbf{#1}}}%
\newcommand{\hlkwb}[1]{\textcolor[rgb]{0.69,0.353,0.396}{#1}}%
\newcommand{\hlkwc}[1]{\textcolor[rgb]{0.333,0.667,0.333}{#1}}%
\newcommand{\hlkwd}[1]{\textcolor[rgb]{0.737,0.353,0.396}{\textbf{#1}}}%

\usepackage{framed}
\makeatletter
\newenvironment{kframe}{%
 \def\at@end@of@kframe{}%
 \ifinner\ifhmode%
  \def\at@end@of@kframe{\end{minipage}}%
  \begin{minipage}{\columnwidth}%
 \fi\fi%
 \def\FrameCommand##1{\hskip\@totalleftmargin \hskip-\fboxsep
 \colorbox{shadecolor}{##1}\hskip-\fboxsep
     \hskip-\linewidth \hskip-\@totalleftmargin \hskip\columnwidth}%
 \MakeFramed {\advance\hsize-\width
   \@totalleftmargin\z@ \linewidth\hsize
   \@setminipage}}%
 {\par\unskip\endMakeFramed%
 \at@end@of@kframe}
\makeatother

\definecolor{shadecolor}{rgb}{.97, .97, .97}
\definecolor{messagecolor}{rgb}{0, 0, 0}
\definecolor{warningcolor}{rgb}{1, 0, 1}
\definecolor{errorcolor}{rgb}{1, 0, 0}
\newenvironment{knitrout}{}{} % an empty environment to be redefined in TeX

\usepackage{alltt}

\usepackage{thumbpdf,lmodern}

\usepackage{bm}  % bold math
\usepackage{amsmath,amssymb}  % AMS environments

\usepackage{xparse}  % smarter commands

\newcommand{\fct}[1]{\code{#1()}}

\newcommand{\dimy}{m}
\newcommand{\leny}{n}
\newcommand{\nfac}{r}
\newcommand{\nreg}{K}

\newcommand{\stochvol}{\pkg{stochvol}}
\newcommand{\factorstochvol}{\pkg{factorstochvol}}

\def\yvec{\bm{y}}
\def\fvec{\bm{f}}
\def\hvec{\bm{h}}

\newcommand{\phipar}{\varphi}
\newcommand{\mupar}{\mu}
\newcommand{\sigmapar}{\sigma}
\newcommand{\rhopar}{\rho}
\newcommand{\nupar}{\nu}
\newcommand{\hpar}{h}
\NewDocumentCommand{\Sigmay}{o}{\bm{\IfValueTF{#1}{#1}{}\Sigma}}  % usage: \Sigmay or \Sigmay[\tilde]  (tilde is also bold face then!)
\newcommand{\deffsvpars}[2]{%
  \expandafter\def\csname phi#1\endcsname{#2{\phipar}}
  \expandafter\def\csname mu#1\endcsname{#2{\mupar}}
  \expandafter\def\csname sigma#1\endcsname{#2{\sigmapar}}
  \expandafter\def\csname rho#1\endcsname{#2{\rhopar}}
  \expandafter\def\csname h#1\endcsname{#2{\hpar}}
  \expandafter\def\csname Sigma#1\endcsname{\Sigmay[#2]}
}
\deffsvpars{idi}{\bar}   % define all at once
\deffsvpars{fac}{\tilde} % define all at once
\newcommand{\betapriorf}{\bm{b_\beta}}
\newcommand{\betapriors}{\bm{B_\beta}}
\newcommand{\mupriorf}{b_\mupar}
\newcommand{\mupriors}{B_\mupar}
\newcommand{\phipriorf}{a_\phipar}
\newcommand{\phipriors}{b_\phipar}
\newcommand{\rhopriorf}{a_\rhopar}
\newcommand{\rhopriors}{b_\rhopar}
\newcommand{\nuprior}{\lambda_\nupar}
\newcommand{\sigmaprior}{B_\sigmapar}

\newcommand{\Sigmalow}{\Sigmay[\check]}
\newcommand{\Loading}{\Lambda}
\newcommand{\Loadings}{\bm{\Lambda}}

\newcommand{\Real}{\mathbb{R}}

\newcommand{\Student}[3]{t_{#1}(#2,#3)}   % \Student{df}{mean}{scale}
\newcommand{\Exponential}[1]{\mathcal{E}(#1)}   % \Exponential{a}
\newcommand{\Betadist}[2]{\mathcal{B}(#1,#2)}   % \Betadist{a}{b}
\newcommand{\Gammadist}[2]{\mathcal{G}(#1,#2)}   % \Gammadist{a}{b}
\NewDocumentCommand{\Normal}{o o m m}{\mathcal{N}\IfValueTF{#1}{_{#1}}{}\IfValueTF{#2}{\left}{}({#3},{#4}\IfValueTF{#2}{\right}{})}
\newcommand{\Permmat}{\bm{P}}

\DeclareMathOperator{\rank}{rank}
\DeclareMathOperator{\diag}{diag}

\def\svpars{\bm{\vartheta}}
\def\svlpars{\bm{\zeta}}

\graphicspath{{Figures/}}

\author{Darjus Hosszejni\\[.3em]WU Vienna University\\[-.1em]of Economics and Business
\And Gregor Kastner\\[.3em]University of Klagenfurt}
\Plainauthor{Darjus Hosszejni, Gregor Kastner}

\title{Modeling Univariate and Multivariate Stochastic Volatility in \proglang{R} with \stochvol{} and \factorstochvol}
\Plaintitle{Modeling Univariate and Multivariate Stochastic Volatility in R with stochvol and factorstochvol}
\Shorttitle{Modeling Stochastic Volatility in \proglang{R}}

\Abstract{
Stochastic volatility (SV) models are nonlinear state-space models that enjoy increasing popularity for fitting and predicting heteroskedastic time series.
However, due to the large number of latent quantities, their efficient estimation is non-trivial and software that allows to easily fit SV models to data is rare.
  We aim to alleviate this issue by presenting novel implementations of five SV models delivered in two \proglang{R} packages.
  Several unique features are included and documented.
  As opposed to previous versions, \stochvol{} is now capable of handling linear mean models, conditionally heavy tails, and the leverage effect in combination with SV.
  Moreover, we newly introduce \factorstochvol{} which caters for multivariate SV.
  Both packages offer a user-friendly interface through the conventional \proglang{R} generics and a range of tailor-made methods.
  Computational efficiency is achieved via interfacing \proglang{R} to \proglang{C++} and doing the heavy work in the latter.
  In the paper at hand, we provide a detailed discussion on Bayesian SV estimation and showcase the use of the new software through various examples.
}

\Plainkeywords{%
Bayesian inference,
state-space model,
heteroskedasticity,
dynamic correlation,
dynamic covariance,
factor stochastic volatility,
Markov chain Monte Carlo (MCMC),
leverage effect,
asymmetric return distribution,
heavy tails,
financial time series%
}

\Address{
  Darjus Hosszejni\\
  Institute for Statistics and Mathematics\\
  Department of Finance, Accounting and Statistics\\
  WU Vienna University of Economics and Business\\
  Welthandelsplatz~1 / Building D4 / Level 4\\
  1020 Vienna, Austria\\
  E-mail: \email{Darjus.Hosszejni@wu.ac.at}\\
  URL: \url{http://statmath.wu.ac.at/~hosszejni/}
  \\\\
  Gregor Kastner\\
  Department of Statistics\\
  University of Klagenfurt\\
  Universit\"atsstraße 65-67\\
  9020 Klagenfurt, Austria\\
  E-mail: \email{Gregor.Kastner@aau.at}\\
  URL: \url{http://statmath.wu.ac.at/~kastner/}
}
\IfFileExists{upquote.sty}{\usepackage{upquote}}{}
\begin{document}

\section*{Preface}
This preprint corresponds to the forthcoming article of the same name~\citep{thisjss}.

\section[Introduction]{Introduction} \label{sec:intro}

Time dependent variance is an indispensable ingredient of financial and economic time series modeling.
Already~\citet{markowitz1952portfolio} concerns himself with methods that take into account heteroskedasticity in a better way than a rolling window estimation.
By 1982, two fundamentally different approaches had been developed to cater to these needs.
On the one hand, \citet{engle1982arch} lays the groundwork for a family of time varying volatility models, most notably the generalized autoregressive conditional heteroskedasticity model~\citep[GARCH,][]{bollerslev1986garch}.
These models feature conditionally deterministic changes in the variance.
\citet{taylor1982sv}, on the other hand, addresses heteroskedasticity in his seminal work with a non-linear latent state space model, later coined the stochastic volatility (SV) model.
There, the volatility process evolves in a stochastic manner.
Despite some empirical evidence in favor of SV models over their corresponding GARCH counterparts~\citep{jacquier1994bayesian,ghysels1996stochastic,kim1998stochastic,nakajima2012bayesian}, SV and its variants enjoy little publicity among practitioners.
As~\citet{bos2012relating} underlines, one reason for this might be the lack of standard software.
In response, \citet{kastner2016dealing} provides a first version of the \proglang{R}~\citep{rlanguage} package \stochvol{} but fails to feature conditional non-Gaussianity, asymmetry (the so-called leverage effect), and multivariate generalizations.

We address these shortcomings in the manuscript at hand.
First, we extend \stochvol{}~\citep{kastner2020stochvol} with several practically relevant univariate methods.
Second, we introduce the new companion package \factorstochvol~\citep{factorstochvol} which focuses on the multivariate case.
The extended \stochvol{} now provides the means for the Bayesian estimation of vanilla SV, heavy-tailed SV, SV with leverage, and heavy-tailed SV with leverage~\citep{harvey1996estimation,omori2007stochastic,nakajima2012stochastic}.
Moreover, the package also handles these models naturally when embedded into a linear model or an autoregressive (AR) context.
The \factorstochvol{} package implements an efficient method for the Bayesian estimation of the factor SV model~\citep{kastner2017efficient}.
Among other features, the package provides several automatic factor identification schemes, hierarchical shrinkage priors~\citep[variations of the normal gamma prior,][]{griffin2010inference}, and an array of intuitive visualization methods for the high-dimensional posteriors.

The remainder of this paper is structured as follows.
We formally introduce the univariate and the multivariate models in Sections~\ref{sec:svmodels} and~\ref{sec:fsv}, respectively, including a discussion about prior distributions and a brief overview of the estimation methods.
In Section~\ref{sec:stochvol}, we unveil the new samplers of the \stochvol{} package through three example models.
We describe the \factorstochvol{} package in Section~\ref{sec:factorstochvol}, and then we conclude.

\section{Univariate SV models}\label{sec:svmodels}

We begin by introducing the vanilla SV model with linear regressors, henceforth simply called the SV model.
This is a minor but important extension of the SV model without regressors.
We also settle the notation and establish a baseline model that we generalize and reuse throughout the manuscript.
Consequently, we proceed with three generalized models: the SV model with Student's $t$~errors (SVt), the SV model with leverage (SVl), and their combination, the SV model with Student's $t$~errors and leverage (SVtl).
Finally, we close the section after discussing prior distributions and Markov chain Monte Carlo (MCMC) sampling.

\subsection{Model specifications}\label{sec:unimodels}

The key feature of the SV model is its stochastic and time-varying specification of the variance evolution.
In particular, the log-variance is assumed to follow an AR(1) process.
This feature unites the following models.

\subsubsection{Vanilla SV with linear regressors}

Let $\yvec=(y_1,\dots,y_\leny)^\top$ denote a vector of observations.
The SV model assumes the following structure for $\yvec$,
\begin{equation}\label{eq:vanillasv}
  \begin{split}
    y_t &= \bm x_t\bm\beta + \exp(\hpar_t/2)\varepsilon_t, \\
    \hpar_{t+1} &= \mupar+\phipar(\hpar_t-\mupar) + \sigmapar\eta_t, \\
    \varepsilon_t &\sim \Normal{0}{1}, \\
    \eta_t &\sim \Normal{0}{1},
  \end{split}
\end{equation}
where $\Normal{b}{B}$ denotes the normal distribution with mean $b\in\mathbb{R}$ and variance $B\in\mathbb{R}^+$, and $\varepsilon_t$ and $\eta_t$ are independent.
The log-variance process $\hvec=(\hpar_1,\dots,\hpar_\leny)^\top$ is initialized by $\hpar_0\sim\Normal{\mupar}{\sigmapar^2/(1-\phipar^2)}$.
$\bm X=(\bm x_1^\top,\dots,\bm x_\leny^\top)^\top$ is an $\leny\times\nreg$ matrix containing in its $t$th row the vector of $\nreg$ regressors at time $t$.
The $\nreg$ regression coefficients are collected in $\bm\beta=(\beta_1,\dots,\beta_\nreg)^\top$.
We refer to $\svpars=(\mupar,\phipar,\sigmapar)$ as the SV parameters: $\mupar$ is the level, $\phipar$ is the persistence, and $\sigmapar$ (also called \emph{volvol}) is the standard deviation of the log-variance.

\subsubsection{SV with Student's t errors}

Several authors have suggested to use non-normal conditional residual distributions for stochastic volatility modeling.
Examples include the Student's $t$~distribution \citep{harvey1994multivariate}, the extended generalized inverse Gaussian \citep{silva2006extented}, (semi-)parametric residuals \citep{jensen2010bayesian,delatola2011bayesian}, or the generalized hyperbolic skew Student's $t$~distribution \citep{nakajima2012stochastic}.
We implement Student's $t$~errors for the observation equation in \stochvol{}.
Formally,
\begin{equation}\label{eq:heavysv}
  \begin{split}
    y_t &= \bm x_t\bm\beta + \exp(\hpar_t/2)\varepsilon_t, \\
    \hpar_{t+1} &= \mupar+\phipar(\hpar_t-\mupar) + \sigmapar\eta_t, \\
    \varepsilon_t &\sim \Student{\nupar}{0}{1}, \\
    \eta_t &\sim \Normal{0}{1},
  \end{split}
\end{equation}
where $\varepsilon_t$ and $\eta_t$ are independent.
$\Student{\nupar}{a}{b}$ is the Student's $t$~distribution with $\nupar$ degrees of freedom, mean $a$, and variance $b$.
The single difference between Equation~\ref{eq:vanillasv} and Equation~\ref{eq:heavysv} is that here the observations are conditionally $t$~distributed.
Hence, Equation~\ref{eq:heavysv} generalizes Equation~\ref{eq:vanillasv} through the new parameter $\nupar$ as the Student's $t$~distribution converges in law to the standard normal distribution when $\nupar$ goes to infinity.

\subsubsection{SV with leverage}

Propositions for asymmetric innovations include non-parametric distributions~\citep{jensen2014estimating}, skewed distributions~\citep{nakajima2012stochastic}, and distributions featuring correlation with the variance process, also called the leverage effect~\citep{harvey1996estimation,jacquier2004bayesian}.
We implement the leverage effect in the \stochvol{} package.
Formally,
\begin{equation}\label{eq:svleverage}
  \begin{split}
    y_t &= \bm x_t\bm\beta + \exp(\hpar_t/2)\varepsilon_t, \\
    \hpar_{t+1} &= \mupar+\phipar(\hpar_t-\mupar) + \sigmapar\eta_t, \\
    \varepsilon_t &\sim \Normal{0}{1}, \\
    \eta_t &\sim \Normal{0}{1},
  \end{split}
\end{equation}
where the correlation matrix of $(\varepsilon_t,\eta_t)$ is
\begin{equation}\label{eq:corrmatrix}
  \bm\Sigma^\rhopar = \begin{pmatrix} 1 && \rhopar \\ \rhopar && 1 \end{pmatrix}.
\end{equation}
The vector $\svlpars=(\mupar,\phipar,\sigmapar,\rhopar)^\top$ collects the SV parameters.
The new parameter compared to Equation~\ref{eq:vanillasv} is a correlation term $\rhopar$ which relates the residuals of the observations to the innovations of the variance process.
Equation~\ref{eq:vanillasv} is therefore a special case of Equation~\ref{eq:svleverage} with a pre-fixed $\rhopar=0$.

\subsubsection{SV with Student's t errors and leverage}

Some authors have proposed the combination of $t$~errors with the leverage effect~\citep{jacquier2004bayesian,omori2007stochastic,nakajima2009leverage}.
We implement the common generalization of Equation~\ref{eq:heavysv} and Equation~\ref{eq:svleverage}.
Formally,
\begin{equation}\label{eq:heavysvleverage}
  \begin{split}
    y_t &= \bm x_t\bm\beta + \exp(\hpar_t/2)\varepsilon_t, \\
    \hpar_{t+1} &= \mupar+\phipar(\hpar_t-\mupar) + \sigmapar\eta_t, \\
    \varepsilon_t &\sim \Student{\nupar}{0}{1}, \\
    \eta_t &\sim \Normal{0}{1},
  \end{split}
\end{equation}
where the correlation matrix of $(\varepsilon_t,\eta_t)$ is $\bm\Sigma^\rhopar$ as in Equation~\ref{eq:corrmatrix}.

\subsection{Prior distributions}\label{sec:svpriors}

We a priori assume $\bm\beta\sim\Normal[\nreg]{\betapriorf}{\betapriors}$, where $\Normal[l]{\bm b}{\bm B}$ is the $l$-dimensional normal distribution with mean vector $\bm b$ and variance-covariance matrix $\bm B$.
For small values in the diagonal of $\betapriors$, this prior enforces shrinkage towards $\betapriorf$; for large values in the diagonal, the prior turns rather uninformative.
By setting $\betapriorf$ to the zero vector and $\betapriors$ to a scaled identity matrix, the prior distribution becomes the Bayesian analogue to ridge regression~\citep[see, e.g.,][for a discussion of this and other shrinkage priors]{park2008bayesian}.

The level $\mupar\in\Real$ is unrestricted, hence we can apply the common $\mupar\sim\Normal{\mupriorf}{\mupriors}$ prior.
Depending on the application, a fairly uninformative distribution is the usual choice, e.g., setting $\mupriorf=0$ and $\mupriors\ge100$ for daily asset log returns.
In our experience, the exact values of the prior mean and prior variance of $\mupar$ do not strongly affect the estimation results unless $\mupriors$ is small.

To achieve stationarity in the variance process, a restricted persistence $\phipar\in(-1,1)$ is needed.
To this end, we assume $(\phipar+1)/2\sim\Betadist{\phipriorf}{\phipriors}$, where $\Betadist{\phipriorf}{\phipriors}$ is the beta distribution with shape parameters $\phipriorf$ and $\phipriors$.
The selection of the shape parameters may be relatively influential with many data sets.
In financial applications with daily asset log returns, the variance tends to be highly persistent, i.e., $\phipar\approx1$.
Such domain knowledge can be used as prior information by allocating more probability to positive high values of $\phipar$, e.g., by setting $\phipriorf\gtrsim5$ and $\phipriors\approx1.5$.
As an alternative, when stationarity is not assumed, the untruncated prior $\phipar\sim\Normal{b_\phipar}{B_\phipar}$ can also be applied.

The volvol is positive but we would like allow $\sigmapar$ to approach 0 as closely as needed -- that allows us to be less informative and to improve the estimates.
Following~\citet{fruhwirth2010stochastic} and~\citet{kastner2014ancillarity}, we advocate $\sigmapar\sim\lvert\Normal{0}{\sigmaprior}\rvert$ instead, where $\lvert\Normal{0}{\sigmaprior}\rvert$ denotes the half normal distribution.
It corresponds to $\sigmapar^2\sim\Gammadist{1/2}{1/(2\sigmaprior)}$, where $\Gammadist{a}{b}$ is the gamma distribution with shape parameter $a$ and rate parameter $b$.
As an alternative, the commonly applied and convenient conjugate gamma prior on $\sigmapar^{-2}$ can be assumed.
However, it bounds $\sigmapar$ away from 0 and it is therefore in our view an unsatisfactory choice.

As a last step in fully specifying the vanilla SV model in Equation~\ref{eq:vanillasv}, the variance process is initialized a priori with its stationary distribution, i.e., $\hpar_0\sim\Normal{\mupar}{\sigmapar^2/(1-\phipar^2)}$.
This consistently extends our prior assumptions about $\hvec$ following a stationary AR(1) process.
As an alternative, when stationarity is not assumed, $\hpar_0\sim\Normal{\mupar}{B_h}$ can be applied with a constant variance $B_h$.

The SV models with Student's $t$~errors additionally require the prior specification of the degrees of freedom parameter $\nupar$.
To ascertain interpretability of the scaling $\exp(\hpar_t/2)$, we ensure finite second moments of $\yvec$ by enforcing $\nupar>2$.
As a reviewer recommended, we follow~\citep{geweke1993bayesian} and equip $\nupar$ with an exponential prior $\nupar-2\sim\Exponential{\nuprior}$, where $\nuprior$ is the rate of the exponential distribution.

Finally, in the case of the SV models with leverage, we employ the translated and scaled beta distribution for $\rhopar\in(-1,1)$ as in~\citet{omori2007stochastic}, i.e., $(\rhopar+1)/2\sim\Betadist{\rhopriorf}{\rhopriors}$.
We find that the posterior estimates of $\rhopar$ can be sensitive to its prior distribution, thus, some care is needed when setting the hyperparameters in practice.
In our experience, slightly informative choices such as $\rhopriorf=\rhopriors\approx4$ work well in financial applications.

\subsection{Estimation}

All methods implemented in \stochvol{} and \factorstochvol{} rely on the Bayesian paradigm.\footnote{At this point we would also like to point out works aiming at estimating stochastic volatility and related models within the frequentist framework, see, e.g., \citet{abanto2017maximum, creal2017class}, and in particular the recent \pkg{stochvolTMB} package \citep{stochvolTMB}.}
Bayesian analysis aims to estimate model parameters through Bayesian updating.
By using probability distributions to represent information, Bayes' theorem can be employed to update the prior information to the posterior information by incorporating the observations. This approach has the advantage of providing full uncertainty quantification in a probabilistic framework without relying on asymptotic results; moreover, so-called \emph{shrinkage priors} can be used to regularize the posterior and guard against overfitting. For an introductory textbook on Bayesian statistics, see, for instance, \citet{mcelreath2015statistical}.

When the posterior distribution is not available analytically, one customarily resorts to approximations such as perfect simulation~\citep{huber2015perfect}, approximate Bayesian computation~\citep{sisson2018handbook}, adaptive Monte Carlo methods~\citep{roberts2005coupling}, or MCMC methods.
When computationally feasible, MCMC is a valuable tool that provides draws from the posterior distribution in question.
That way, MCMC approximates the posterior distribution similarly to a histogram approximating a density.
For a more in-depth introduction on MCMC methods, see, for instance, \citet{brooks2011handbook}.

The estimation algorithm of SV, SVt, SVl, and SVtl all resemble the original methodology developed in~\citet{kastner2014ancillarity} for the vanilla SV model.
Namely, to draw from the posterior distribution of $\hvec$ efficiently, the MCMC sampler resorts to approximate mixture representations of Equations~\ref{eq:vanillasv},~\ref{eq:heavysv},~\ref{eq:svleverage} and~\ref{eq:heavysvleverage} similar to the ones in \citet{kim1998stochastic} and \citet{omori2007stochastic}.
Doing so yields a conditionally Gaussian state space model for which efficient sampling methods are available~\citep{fruhwirth1994data,carter1994gibbs}.
Following~\cite{rue2001fast} and~\cite{mccausland2011simulation}, we draw the full vector $\hvec$ ``all without a loop'' (AWOL).

When Student's $t$~errors with unknown degrees of freedom are used, we handle the added complication through the well-known representation of the $t$~distribution as a scale mixture of Gaussians.
This requires additional Gibbs and independence Metropolis-Hastings steps documented in~\citet{kastner2015heavy}.
Furthermore, we deal with the increased complexity in the posterior space of the leverage case by repeated ancillarity-sufficiency interweaving strategies~\citep[ASIS,][]{yu2011to} steps in the sampling scheme, see~\citet{hosszejni2019approaches} for details.

To verify the correctness of the implementation, unit tests are included in the package which can be run by \code{devtools::test()}~\citep{devtools}.
In particular, a variant of Geweke's test~\citep{geweke2004getting} is part of the test suite.
In this test, we exploit that the sampling distribution of the model parameters during the Geweke test is identical to their preset prior distribution.
Therefore, the cumulative distribution function maps the sample to a uniform distribution, which in turn is mapped to a normal distribution using the normal distribution's quantile function.
If the user chooses to execute the automated unit tests in \stochvol{}, the system evaluates the thinned and transformed sample using the \code{shapiro.test()} function, where the thinning of the sample is done to approximate independent sampling.

For maximal computational effectiveness, all sampling algorithms are implemented in the compiled language \proglang{C++}~\citep{cpplanguage} with the help of the \proglang{R} package \pkg{Rcpp}~\citep{eddelbuettel2011rcpp}.
Matrix computations make use of the efficient \proglang{C++} template library \pkg{Armadillo}~\citep{sanderson2016armadillo} through the \proglang{R} package \pkg{RcppArmadillo}~\citep{eddelbuettel2014rcpparmadillo}.\footnote{For explicit run time discussions please see \cite{kastner2014ancillarity} and \cite{hosszejni2019approaches}. For the possibility to use multi-core computing within a single MCMC chain and potential speed gains when doing so, please see \cite{kastner2019sparse}.}
After sampling, the resulting \proglang{R} objects make use of plotting and summary functions of the \proglang{R} package \pkg{coda}~\citep{coda}.

\section{Multivariate SV models} \label{sec:fsv}

A key difficulty accompanying dynamic covariance estimation is the relatively high number of unknowns compared to the number of observations.
More precisely, letting $\dimy$ denote the cross-sectional dimension, the corresponding covariance matrix $\Sigmay_t$ contains $\dimy(\dimy+1)/2$ degrees of freedom, a quadratic term in $\dimy$.
Table~\ref{tab:numpar1} illustrates the ``curse of dimensionality'' for various values of $\dimy$.
One way to break this curse is to use latent factors and thereby achieve a sparse representation of $\Sigmay_t$.

\begin{table}[t]
  \centering
  \begin{tabular}{lll}
    $\dimy$ & free elements of $\Sigmay_t$ & free elements of $\Sigmay_t$ per data point \\ \hline
    1       & 1                            & 1                                           \\
    10      & 55                           & 5.5                                         \\
    100     & 5050                         & 50.5                                        \\
    1000    & 500500                       & 500.5
  \end{tabular}
  \caption{Absolute and relative numbers of free elements of the time-varying covariance matrix $\Sigmay_t$ for different numbers of component series $\dimy$.}
  \label{tab:numpar1}
\end{table}

\subsection{The factor SV model}

Latent factor models embody the idea that even high dimensional systems are driven by only a few sources of randomness.
These few sources of randomness control a few factors, which in turn account for the interactions between the observations.
Moreover, latent factor models provide an efficient tool for dynamic covariance matrix estimation.
They allow for a reduction in the number of unknowns.
A conventional latent factor model with $\nfac$ factors implies the decomposition
\begin{equation}\label{eq:decomp}
  \Sigmay_t=\Sigmalow_t+\Sigmaidi_t,
\end{equation}
where $\rank(\Sigmalow_t)=\nfac<\dimy$, and $\Sigmaidi_t$ is the diagonal matrix containing the variances of the idiosyncratic errors.
The rank assumption on the symmetric $\Sigmalow_t$ gives rise to the factorization $\Sigmalow_t=\Psi\Psi^\top$, where $\Psi\in\Real^{\dimy\times\nfac}$ contains $mr-\nfac(\nfac-1)/2$ free elements~\citep[see, e.g., the pivoted Cholesky algorithm in][]{higham1990analysis}.
Hence, $\dimy(\nfac+1)-\nfac(\nfac-1)/2$ free elements remain in $\Sigmay_t$, now only linear in $\dimy$.
Table~\ref{tab:numpar2} illustrates the ``broken curse of dimensionality'' for various values of $\dimy$ and $\nfac=4$.

\begin{table}[t]
  \centering
  \begin{tabular}{lll}
    $\dimy$ & free elements of $\Sigmay_t$ & free elements of $\Sigmay_t$ per data point \\ \hline
    10      & 44                           & 4.4                                         \\
    100     & 494                          & 4.94                                        \\
    1000    & 4994                         & 4.994
  \end{tabular}
  \caption{Absolute and relative numbers of free elements of the time-varying covariance matrix $\Sigmay_t$ in a factor model for different numbers of component series $\dimy$ and number of factors $\nfac=4$.}
  \label{tab:numpar2}
\end{table}

In the following, we describe the factor SV model employed in the \factorstochvol{} package.
We model the observations $\yvec_t=(y_{t1},\dots,y_{t\dimy})^\top$ as follows.
\begin{equation}\label{eq:fsv}
  \begin{split}
    \yvec_t\mid\bm\beta,\Loadings,\fvec_t,\Sigmaidi_t &\sim \Normal[\dimy]{\bm\beta + \Loadings\fvec_t}{\Sigmaidi_t}, \\
    \fvec_t\mid\Sigmafac_t &\sim \Normal[\nfac]{\bm 0}{\Sigmafac_t},
  \end{split}
\end{equation}
where $\fvec_t=(f_{t1},\dots,f_{t\nfac})^\top$ is the vector of factors, $\bm\beta=(\beta_1,\dots,\beta_\dimy)^\top$ is an observation-specific mean, and $\Loadings\in\Real^{\dimy\times\nfac}$ is a tall matrix holding the factor loadings.
The covariance matrices $\Sigmaidi_t$ and $\Sigmafac_t$ are both diagonal representing independent vanilla SV processes.
\begin{equation}\label{eq:fsvcov}
  \begin{split}
    \Sigmaidi_t &= \diag(\exp(\hidi_{t1}),\dots,\exp(\hidi_{t\dimy})), \\
    \Sigmafac_t &= \diag(\exp(\hfac_{t1},\dots,\exp(\hfac_{t\nfac}))), \\
    \hidi_{ti} &\sim \Normal{\muidi_i+\phiidi_i(\hidi_{t-1,i}-\muidi_i)}{\sigmaidi_i^2}, \quad i=1,\dots,\dimy, \\
    \hfac_{tj} &\sim \Normal{\mufac_j+\phifac_j(\hfac_{t-1,j}-\mufac_j)}{\sigmafac_j^2}, \quad j=1,\dots,\nfac.
  \end{split}
\end{equation}
For a more theoretical treatment of factor SV from a Bayesian point of view, the reader is referred to, e.g., \citet{pitt1999time}, \citet{aguilar2000bayesian}, \citet{chib2006analysis}, and \citet{han2006asset}.

Based on Equation~\ref{eq:fsv}, we can reformulate Equation~\ref{eq:decomp} as
\begin{equation}\label{eq:vardecomp}
  \Sigmay_t=\Loadings\Sigmafac_t\Loadings^\top+\Sigmaidi_t,
\end{equation}
from which several identification issues are apparent: the order, the sign, and the scale of the factors is unidentified.
More specifically, for any generalized permutation matrix%
\footnote{A generalized permutation matrix has the zero--non-zero pattern of a permutation matrix, but it is allowed to have any non-zero values instead of just ones.
Hence, a generalized permutation matrix not only permutes but also scales and switches the sign of its multiplier.} %
$\Permmat$ of size $\nfac\times\nfac$, we find another valid decomposition $\Sigmay_t=\Loadings^\prime\Sigmafac^\prime_t(\Loadings^\prime)^\top+\Sigmaidi_t$, where $\Loadings^\prime=\Loadings\Permmat^{-1}$ and
$\Sigmafac^\prime_t=\Permmat\Sigmafac_t\Permmat^\top$.
We resolve the ambiguity in the scale of the factors by fixing the level of their log-variance to zero, i.e., $\mufac_j=0$ for $j=1,\dots,\nfac$.
Sign and order identification can be enforced through restrictions on the factor loadings matrix $\Loadings$.
Several options are available in \factorstochvol{} for restricting $\Loadings$, for details see Section~\ref{sec:fsvuse2}.

\subsection{Prior distributions}\label{sec:fsvpriors}

Priors need to be specified for the mean, the latent log-variance processes, and for the factor loadings matrix $\Loadings$.
We choose $\beta_j\sim\Normal{b_\beta}{B_\beta}$, independently for $j=1,\dots,\dimy$.
For small values of $B_\beta$, this shrinks $\beta_j$ toward $b_\beta$; for large values of $B_\beta$, the prior is fairly uninformative.

The log-variance processes have the same prior specification as in the univariate case in Section~\ref{sec:svpriors}.
For $\Loadings$, three types of priors are currently implemented in \factorstochvol.
All three can be written in the form $\Loading_{ij}\sim\Normal{0}{\tau^2_{ij}}$ independently for each applicable $i\in\{1,\dots\dimy\}$ and $j\in\{1,\dots,\nfac\}$.
First, one can fix all the $\tau^2_{ij}$s -- not necessarily to the same value -- a priori.
This results in a normal prior for each element of the loadings matrix.

The second type is a hierarchical prior which has been developed to induce more flexible and potentially stronger shrinkage,
\begin{equation} \label{eq:ngprior}
  \Loading_{ij}\mid\tau_{ij}^2 \sim \Normal{0}{\tau_{ij}^2}, \quad \tau_{ij}^2\mid\lambda_i^2 \sim \Gammadist{a}{a\lambda_i^2/2}. 
\end{equation}
This distribution is termed normal gamma prior by \citet{griffin2010inference} and implies a conditional variance $\mathbb{V}(\Loading_{ij}\mid\lambda_i^2)$ of $2/\lambda_i^2$ and an unconditional excess kurtosis of $3/a$.
The value of $a$ is treated as a structural parameter to be fixed by the user, where choosing $a$ small ($\lesssim 1$) enforces strong shrinkage towards zero, while choosing $a$ large ($\gtrsim 1$) imposes little shrinkage.
The case $a = 1$ is a special case termed the Bayesian Lasso prior~\citep{park2008bayesian}.
The parameter $\lambda_i^2$ is estimated from the data with $\lambda_i^2 \sim \Gammadist{c}{d}$.

The third type is a slight modification of the second.
Because variances in each row of the factor loadings matrix $\Loadings$ can be seen as ``random effects'' from the same underlying distribution, the prior in Equation~\ref{eq:ngprior} induces row-wise shrinkage with element-wise adaption.
Analogously, one could also consider column-wise shrinkage with element-wise adaption, i.e.,
\begin{equation} \label{eq:ngpriorCW}
  \Loading_{ij}\mid\tau_{ij}^2\sim\Normal{0}{\tau_{ij}^2}, \quad \tau_{ij}^2\mid\lambda_j^2\sim\Gammadist{a}{a\lambda_j^2/2},
\end{equation}
with the corresponding prior $\lambda_j^2 \sim \Gammadist{c}{d}$.

\subsection{Estimation}

Bayesian estimation in the factor SV model builds on the univariate vanilla SV implementations in \stochvol{} and features several levels of efficiency boosting. To alleviate the problem of potentially slow convergence in high dimensions, it is carried out via a sampler that utilizes several variants of ASIS. The sampling details implemented in \factorstochvol{} are described in \citet[using Gaussian priors for the factor loadings]{kastner2017efficient} as well as \citet[using hierarchical shrinkage priors for the factor loadings]{kastner2019sparse}.

Similarly to \stochvol{} and in an attempt to make computation time bearable even in higher dimensions, \factorstochvol{}'s main sampler is written in \proglang{C++}.
It uses the \proglang{R} package \pkg{Rcpp} to ease communication between \proglang{R} and \proglang{C++}. The univariate SV parts are borrowed from \stochvol{} and interfaced through its \proglang{C}/\proglang{C++}-level updating function \fct{update\_fast\_sv}. In doing so, moving between interpreted \proglang{R} code and compiled \proglang{C++} code at each MCMC iteration is avoided.

\section[The stochvol package]{The \stochvol{} package} \label{sec:stochvol}%

The \stochvol{} package provides means for fitting univariate SV, SVt, SVl, and SVtl models via its sampling routines \fct{svsample}, \fct{svtsample}, \fct{svlsample}, and \fct{svtlsample}, respectively.
In the following, we describe a recommended workflow with \stochvol.
First, we discuss estimation, visualization, and prediction using default settings.
Then, we show how to adapt the values of the prior hyperparameters and how to configure the sampling mechanism.

\subsection{Preparing the data and running the MCMC sampler} \label{sec:svuse1}

We estimate three models that exemplify the features and the user interface of \stochvol.
Using the \code{exrates} data found in the package, we model the EURCHF exchange rate (the price of 1 euro in Swiss franc) in the period between March 1, 2008 and March 1, 2012 (1028 data points) in three different ways.

\subsubsection{AR(1) model with SV residuals}

The first example is an AR(1) model with SV residuals, i.e., Equation~\ref{eq:vanillasv} turns into
\begin{equation*}
  \begin{split}
    y_t\mid y_{t-1},\beta_0,\beta_1,\hpar_t &\sim \Normal{\beta_0+\beta_1y_{t-1}}{\exp(\hpar_t)}, \\
    \hpar_{t+1}\mid\svpars,\hpar_t &\sim \Normal{\mupar+\phipar(\hpar_t-\mupar)}{\sigmapar^2}.
  \end{split}
\end{equation*}
Using this model, we test whether the exchange rate follows a random walk with SV.
In this case, we expect the posteriors of $\beta_0$ and $\beta_1$ to concentrate around 0 and 1, respectively.

In order to estimate this AR(1)-SV model, we need to prepare the input $\yvec$ as a \code{numeric} sequence of length $\leny$ and pass it as the first input argument to \fct{svsample} as follows:

\begin{knitrout}
\definecolor{shadecolor}{rgb}{0.969, 0.969, 0.969}\color{fgcolor}\begin{kframe}
\begin{alltt}
\hlstd{R> }\hlkwd{set.seed}\hlstd{(}\hlnum{1}\hlstd{)}
\hlstd{R> }\hlkwd{library}\hlstd{(}\hlstr{"stochvol"}\hlstd{)}
\hlstd{R> }\hlkwd{data}\hlstd{(}\hlstr{"exrates"}\hlstd{)}
\hlstd{R> }\hlstd{ind} \hlkwb{<-} \hlkwd{which}\hlstd{(exrates}\hlopt{$}\hlstd{date} \hlopt{>=} \hlkwd{as.Date}\hlstd{(}\hlstr{"2008-03-01"}\hlstd{)} \hlopt{&}
\hlstd{+  }  \hlstd{exrates}\hlopt{$}\hlstd{date} \hlopt{<=} \hlkwd{as.Date}\hlstd{(}\hlstr{"2012-03-01"}\hlstd{))}
\hlstd{R> }\hlstd{CHF_price} \hlkwb{<-} \hlstd{exrates}\hlopt{$}\hlstd{CHF[ind]}
\hlstd{R> }\hlstd{res_sv} \hlkwb{<-} \hlkwd{svsample}\hlstd{(CHF_price,} \hlkwc{designmatrix} \hlstd{=} \hlstr{"ar1"}\hlstd{)}
\end{alltt}
\end{kframe}
\end{knitrout}
We set \code{designmatrix = "ar1"} to use the AR(1) specification.
More generally, \code{designmatrix} can take \code{character} values of the form \code{"ar0"} for a constant mean model, or \code{"ar1"}, \code{"ar2"}, etc., for AR(1), AR(2), and so on.

\subsubsection{Constant mean model with SVt residuals}

The second example is a constant mean model with SVt residuals, i.e., Equation~\ref{eq:heavysv} becomes
\begin{equation*}
  \begin{split}
    y_t\mid\beta_0,\hpar_t,\nupar &\sim \Student{\nu}{\beta_0}{\exp(\hpar_t/2)}, \\
    \hpar_{t+1}\mid\svpars,\hpar_t &\sim \Normal{\mupar+\phipar(\hpar_t-\mupar)}{\sigmapar^2}.
  \end{split}
\end{equation*}
If the returns are heavy-tailed, most of the posterior mass of $\nupar$ concentrates on low values, e.g., smaller than 20.
Otherwise, there is little evidence for high kurtosis.

We compute the log returns by applying \fct{logret} on the previously calculated \code{CHF\_price}.
Then, to estimate the constant mean model with heavy tailed SV residuals, we pass the vector of log returns to \fct{svtsample} with \code{designmatrix} set to \code{"ar0"}.

\begin{knitrout}
\definecolor{shadecolor}{rgb}{0.969, 0.969, 0.969}\color{fgcolor}\begin{kframe}
\begin{alltt}
\hlstd{R> }\hlkwd{set.seed}\hlstd{(}\hlnum{2}\hlstd{)}
\hlstd{R> }\hlstd{CHF_logret} \hlkwb{<-} \hlnum{100} \hlopt{*} \hlkwd{logret}\hlstd{(CHF_price)}
\hlstd{R> }\hlstd{res_svt} \hlkwb{<-} \hlkwd{svtsample}\hlstd{(CHF_logret,} \hlkwc{designmatrix} \hlstd{=} \hlstr{"ar0"}\hlstd{)}
\end{alltt}
\end{kframe}
\end{knitrout}

\subsubsection{Multiple regression with SVl residuals}

The third example is a multiple regression model with an intercept, two regressors, and SVl residuals; that is, Equation~\ref{eq:svleverage} turns into
\begin{equation*}
  \begin{split}
    \begin{pmatrix} y_t \\ \hpar_{t+1} \end{pmatrix} \;\Bigg|\; \hpar_t,\svlpars,\begin{pmatrix} x_{t1} \\ x_{t2} \end{pmatrix},\beta_0,\begin{pmatrix} \beta_1 \\ \beta_2 \end{pmatrix} &\sim
    \Normal[2][1]{\begin{pmatrix} \beta_0+\beta_1x_{t1}+\beta_2x_{t2} \\ \mupar+\phipar(\hpar_t-\mupar) \end{pmatrix}}{\bm\Sigma^\rhopar}, \\
    \bm\Sigma^\rhopar &= \begin{pmatrix} \exp(\hpar_t) && \rhopar\sigmapar\exp(\hpar_t/2) \\ \rhopar\sigmapar\exp(\hpar_t/2) && \sigmapar^2 \end{pmatrix}.
  \end{split}
\end{equation*}
For illustration, we regress EURCHF log returns onto the contemporaneous log returns on EURUSD and EURJPY, the value of 1 euro per US dollar and Japanese yen, respectively.

To estimate a multiple regression model using \stochvol, we need to prepare a \code{numeric} matrix $\bm X$ of dimension $\leny\times\nreg$, where rows correspond to time points and columns to covariates.
We create an intercept as the first column of $\bm X$, and we set the second and the third columns to the EURUSD log returns and the EURJPY log returns, respectively; finally, we use the columns of $\bm X$ as covariates in the multiple regression.
\begin{knitrout}
\definecolor{shadecolor}{rgb}{0.969, 0.969, 0.969}\color{fgcolor}\begin{kframe}
\begin{alltt}
\hlstd{R> }\hlkwd{set.seed}\hlstd{(}\hlnum{3}\hlstd{)}
\hlstd{R> }\hlstd{X} \hlkwb{<-} \hlkwd{cbind}\hlstd{(}\hlkwc{constant} \hlstd{=} \hlnum{1}\hlstd{,} \hlnum{100} \hlopt{*} \hlkwd{logret}\hlstd{(exrates}\hlopt{$}\hlstd{USD[ind]),}
\hlstd{+  }  \hlnum{100} \hlopt{*} \hlkwd{logret}\hlstd{(exrates}\hlopt{$}\hlstd{JPY[ind]))}
\hlstd{R> }\hlstd{res_svl} \hlkwb{<-} \hlkwd{svlsample}\hlstd{(CHF_logret,} \hlkwc{designmatrix} \hlstd{= X)}
\end{alltt}
\end{kframe}
\end{knitrout}

\subsection{Visualizing the results} \label{sec:svuse2}

Often, the joint posterior distribution of model parameters and latent quantities mark the goal of a Bayesian analysis.
To inspect it, one can look at summary statistics and various types of visualizations of marginal posterior distributions.
Also, it is recommended to examine the Markov chain for possible convergence issues -- this happens usually by investigating trace plots of posterior quantities.
For this reason, inspired by the \pkg{coda} package, \stochvol{} provides its own instances of the \proglang{R} generic functions \fct{plot} and \fct{summary}.
In order to introduce the tools that \stochvol{} provides for analyzing MCMC output, we briefly examine the results of the third example (multiple regression with SVl errors) in the remaining part of the section.

\begin{figure}[t]
  \centering
  \includegraphics[width=\textwidth]{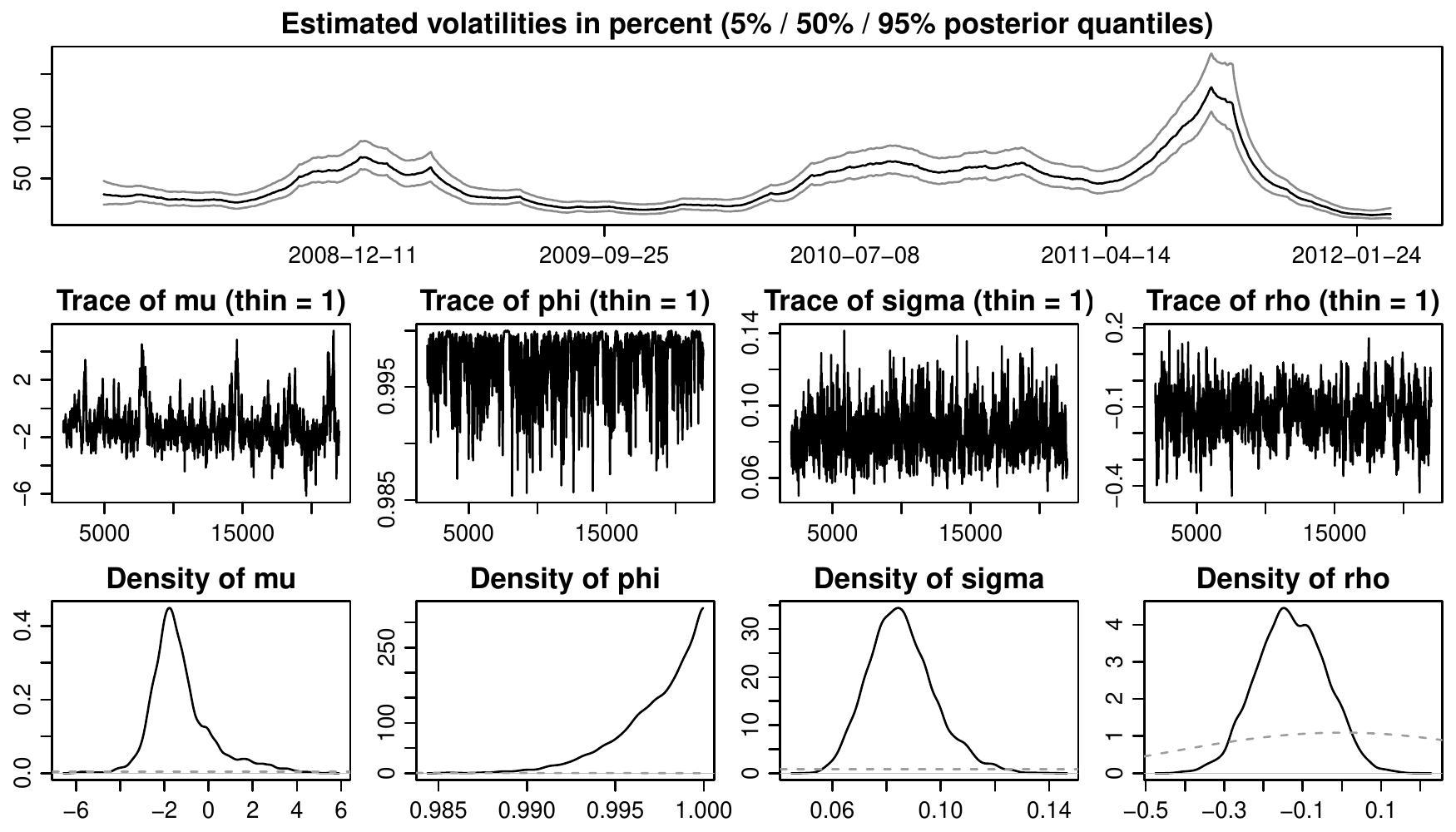}
  \caption{The default plot of an estimated model.
  The top row shows a summary of the posterior of the daily volatility (in percent) $100\exp(\bm h/2)$ through its median (black) and 5\% and 95\% quantiles (gray).
  The remaining panels summarize the Markov chains of the parameters $\mupar$, $\phipar$, $\sigmapar$, and $\rhopar$.
  In particular, the middle row presents trace plots and the bottom row shows prior (gray, dashed) and posterior (black, solid) densities.}
  \label{fig:svlall}
\end{figure}

First, we plot the output of the estimation.
\begin{knitrout}
\definecolor{shadecolor}{rgb}{0.969, 0.969, 0.969}\color{fgcolor}\begin{kframe}
\begin{alltt}
\hlstd{R> }\hlkwd{plot}\hlstd{(res_svl,} \hlkwc{showobs} \hlstd{=} \hlnum{FALSE}\hlstd{,} \hlkwc{dates} \hlstd{= exrates}\hlopt{$}\hlstd{date[ind[}\hlopt{-}\hlnum{1}\hlstd{]])}
\end{alltt}
\end{kframe}
\end{knitrout}
The result is shown in Figure~\ref{fig:svlall}.
We see in the first row a summary of the posterior density of the volatility. Apart from its median, we also receive a quantification of the uncertainty through the 5\% and the 95\% quantiles at each time point.
In the second row, we can follow the evolution of the Markov chain of the SV parameters. In this example, they are $\mupar$, $\phipar$, $\sigmapar$, and $\rhopar$.
Lastly, we see prior and posterior density plots of the parameters in the third row in gray and black, respectively.
They show high persistence and significant leverage.

\begin{figure}[t]
  \centering
  \includegraphics[width=\textwidth]{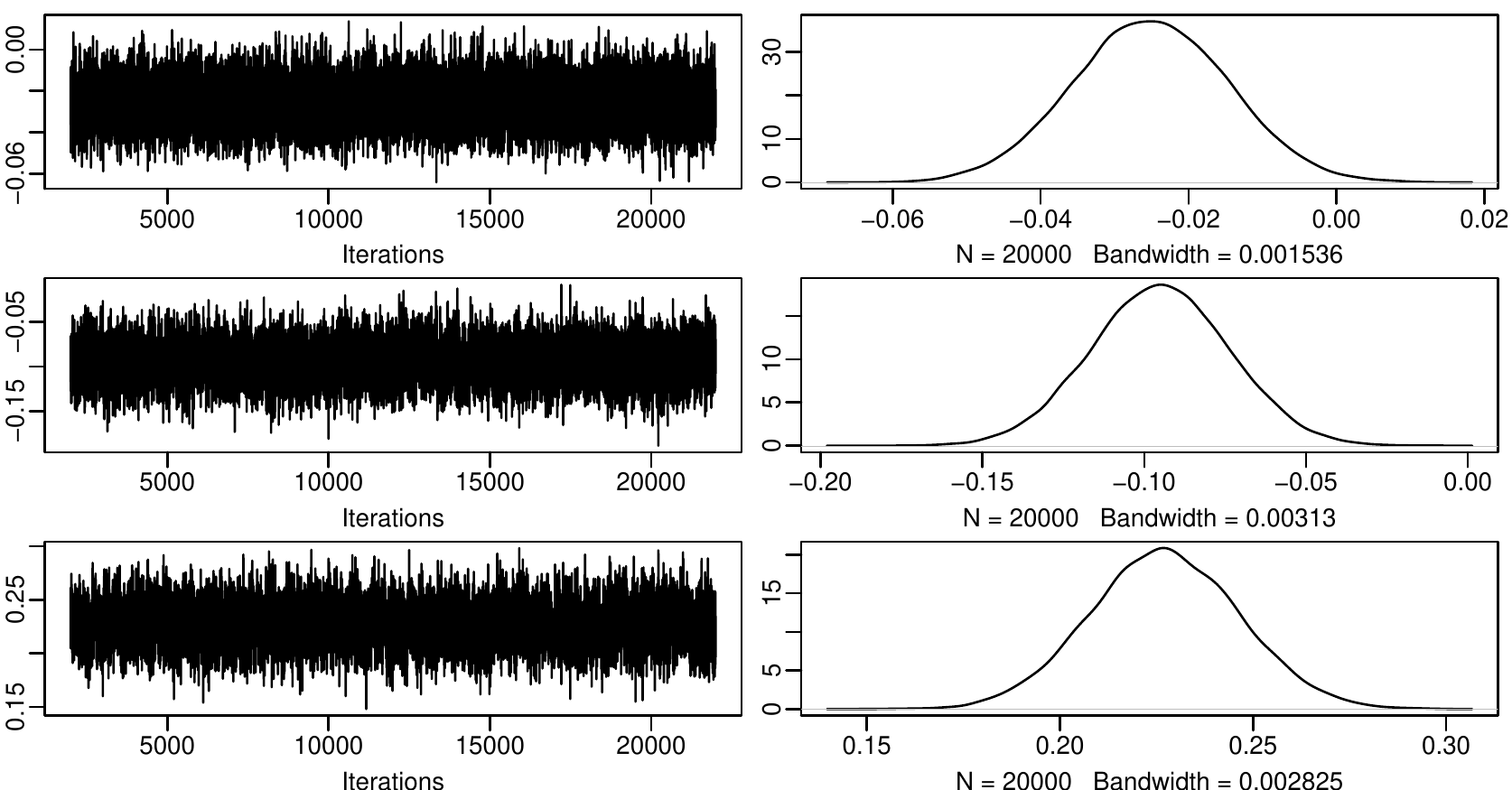}
  \caption{Trace plots and estimated kernel densities of posterior draws from $p(\bm\beta\mid\bm y)$.}
  \label{fig:svlbeta}
\end{figure}

Next, we observe the AR coefficients.
\begin{knitrout}
\definecolor{shadecolor}{rgb}{0.969, 0.969, 0.969}\color{fgcolor}\begin{kframe}
\begin{alltt}
\hlstd{R> }\hlkwa{for} \hlstd{(i} \hlkwa{in} \hlkwd{seq_len}\hlstd{(}\hlnum{3}\hlstd{)) \{}
\hlstd{+  }  \hlstd{coda}\hlopt{::}\hlkwd{traceplot}\hlstd{(}\hlkwd{svbeta}\hlstd{(res_svl)[, i])}
\hlstd{+  }  \hlstd{coda}\hlopt{::}\hlkwd{densplot}\hlstd{(}\hlkwd{svbeta}\hlstd{(res_svl)[, i],} \hlkwc{show.obs} \hlstd{=} \hlnum{FALSE}\hlstd{)}
\hlstd{+  }\hlstd{\}}
\end{alltt}
\end{kframe}
\end{knitrout}
The result is shown in Figure~\ref{fig:svlbeta}.
On the left hand side, we do not spot any signs of convergence or mixing problems in the trace plots.
On the right hand side, we see that none of the posterior densities of $\beta_0$, $\beta_1$, and $\beta_2$ concentrate around 0, hence the covariates seem to have an impact on the dependent variable.

As the final step, we print a numeric summary of the estimation results.
\begin{knitrout}
\definecolor{shadecolor}{rgb}{0.969, 0.969, 0.969}\color{fgcolor}\begin{kframe}
\begin{alltt}
\hlstd{R> }\hlkwd{summary}\hlstd{(res_svl,} \hlkwc{showlatent} \hlstd{=} \hlnum{FALSE}\hlstd{)}
\end{alltt}
\begin{verbatim}
## 
## Summary of 'svdraws' object
## 
## Prior distributions:
## mu        ~ Normal(mean = 0, sd = 100)
## (phi+1)/2 ~ Beta(a = 5, b = 1.5)
## sigma^2   ~ Gamma(shape = 0.5, rate = 0.5)
## nu        ~ Infinity
## rho       ~ Beta(a = 4, b = 4)
## beta      ~ MultivariateNormal(...)
## 
## Stored 20000 MCMC draws after a burn-in of 2000.
## No thinning.
## 
## Posterior draws of SV parameters (thinning = 1):
##              mean     sd      5%     50%   95% ESS
## mu        -1.2554 1.4048 -2.9497 -1.5371 1.727 113
## phi        0.9973 0.0023  0.9927  0.9979 1.000 376
## sigma      0.0855 0.0123  0.0668  0.0846 0.108 344
## rho       -0.1290 0.0877 -0.2731 -0.1307 0.014 226
## exp(mu/2)  0.7458 0.9817  0.2288  0.4637 2.372 113
## sigma^2    0.0075 0.0022  0.0045  0.0072 0.012 344
## 
## Posterior draws of regression coefficients (thinning = 1):
##          mean    sd     5%    50%     95%  ESS
## beta_0 -0.025 0.011 -0.042 -0.025 -0.0078 9506
## beta_1 -0.095 0.021 -0.130 -0.095 -0.0599 3703
## beta_2  0.227 0.019  0.195  0.227  0.2589 2452
\end{verbatim}
\end{kframe}
\end{knitrout}
For brevity, we set \code{showlatent = FALSE} in order not to print all the 1027 latent states.
The output shows the length of the burn-in and the number of draws, the prior specification of the parameters, and a concise summary of the marginal posterior distributions of the parameters $\mupar$, $\phipar$, $\sigmapar$, and $\rhopar$, and additionally of the level of the volatility $\exp(\mupar/2)$ and of $\sigmapar^2$, and of the vector of regression coefficients $\bm\beta$.
This posterior summary is a table consisting of columns for the posterior mean and standard deviation, the 5\%, 50\%, and 95\% quantiles.
The user can influence the shown quantiles by passing a sequence of values between 0 and 1 to \fct{svsample}, \fct{svtsample}, \fct{svlsample}, or \fct{svtlsample} via the argument \code{quantiles}.

The last column in the table depicts the so-called effective sample size (ESS), a measure of the quality of a converged MCMC chain.
Formally, ESS of a Markov chain $\mathrm{C}$ is defined through $M/(1+2\sum_{s=1}^\infty\rho^\text{eff}(s))$, where $M$ is the length of $\mathrm{C}$ and $\rho^\text{eff}(s)$ denotes the autocorrelation function for lag $s$ among the elements of $\mathrm{C}$.
In principle, ESS is the sample size of a serially uncorrelated chain bearing the same Monte Carlo error as our (marginal) chain.
Intuitively speaking, this means that ESS is the number of independent and identically distributed draws that were acquired and gives a sense of how well our chain has explored the posterior space.
Higher values of ESS indicate better mixing.

\subsection[Prediction with stochvol]{Prediction with \stochvol}
\begin{figure}[t]
  \centering
  \includegraphics[trim=40 38 28 58, clip, width=\textwidth]{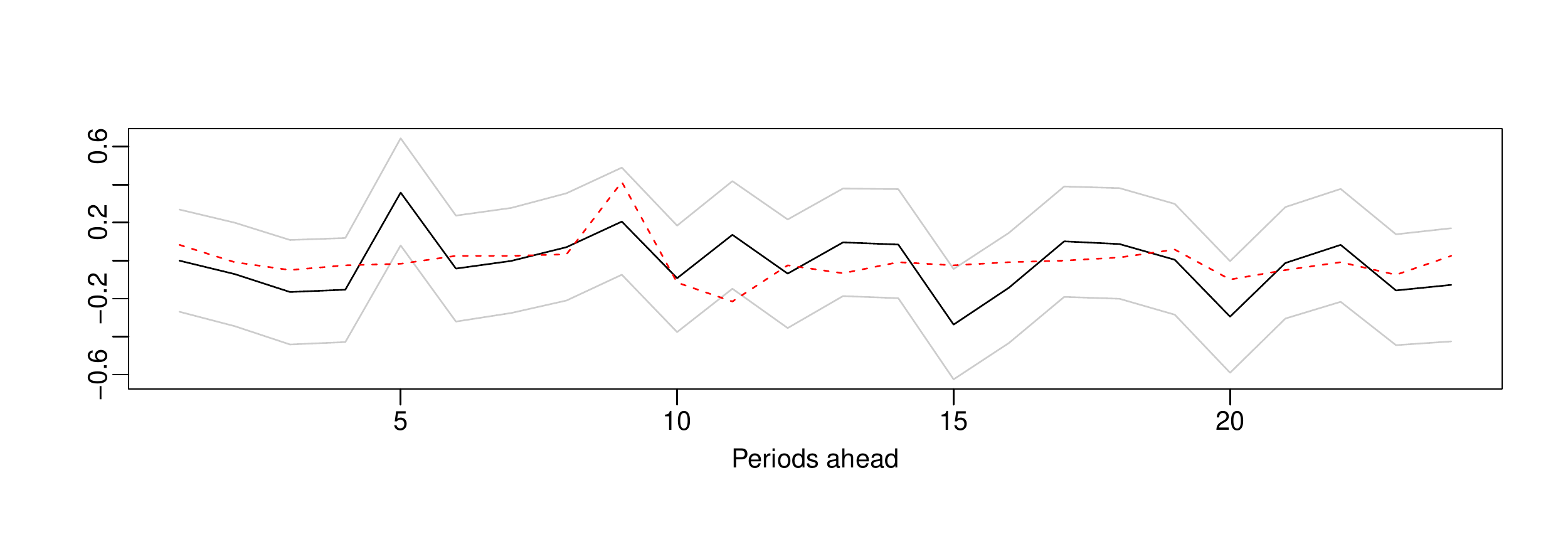}
  \caption{Multi-step ahead predictive distributions (solid, gray and black) and observations (dashed, red).}
  \label{fig:svlpred}
\end{figure}

We employ our estimated model to predict log returns for the remaining days in the data set.
To do so, we first prepare the covariates for the next 24 days and pass them via the argument \code{newdata} of the generic \fct{predict} function along with the estimation output.
Note that we need 25 days of price data to obtain 24 returns.
\begin{knitrout}
\definecolor{shadecolor}{rgb}{0.969, 0.969, 0.969}\color{fgcolor}\begin{kframe}
\begin{alltt}
\hlstd{R> }\hlkwd{set.seed}\hlstd{(}\hlnum{4}\hlstd{)}
\hlstd{R> }\hlstd{pred_ind} \hlkwb{<-} \hlkwd{seq}\hlstd{(}\hlkwd{tail}\hlstd{(ind,} \hlnum{1}\hlstd{),} \hlkwc{length.out} \hlstd{=} \hlnum{25}\hlstd{)}
\hlstd{R> }\hlstd{pred_X} \hlkwb{<-} \hlkwd{cbind}\hlstd{(}\hlkwc{constant} \hlstd{=} \hlnum{1}\hlstd{,} \hlnum{100} \hlopt{*} \hlkwd{logret}\hlstd{(exrates}\hlopt{$}\hlstd{USD[pred_ind]),}
\hlstd{+  }  \hlnum{100} \hlopt{*} \hlkwd{logret}\hlstd{(exrates}\hlopt{$}\hlstd{JPY[pred_ind]))}
\hlstd{R> }\hlstd{pred_svl} \hlkwb{<-} \hlkwd{predict}\hlstd{(res_svl,} \hlnum{24}\hlstd{,} \hlkwc{newdata} \hlstd{= pred_X)}
\end{alltt}
\end{kframe}
\end{knitrout}
As we have access to the entire distribution of future log returns, we can quantify the uncertainty around our predictions through quantiles.
In the following code snippet, we visualize the $k$-step-ahead predictive distributions for $k=1,\dots,24$, along with the truly observed values.
The result is in Figure~\ref{fig:svlpred}.
\begin{knitrout}
\definecolor{shadecolor}{rgb}{0.969, 0.969, 0.969}\color{fgcolor}\begin{kframe}
\begin{alltt}
\hlstd{R> }\hlstd{obs_CHF} \hlkwb{<-} \hlnum{100} \hlopt{*} \hlkwd{logret}\hlstd{(exrates}\hlopt{$}\hlstd{CHF[pred_ind])}
\hlstd{R> }\hlstd{qs} \hlkwb{<-} \hlkwd{t}\hlstd{(}\hlkwd{apply}\hlstd{(}\hlkwd{predy}\hlstd{(pred_svl),} \hlnum{2}\hlstd{, quantile,} \hlkwd{c}\hlstd{(}\hlnum{0.05}\hlstd{,} \hlnum{0.5}\hlstd{,} \hlnum{0.95}\hlstd{)))}
\hlstd{R> }\hlkwd{ts.plot}\hlstd{(}\hlkwd{cbind}\hlstd{(qs, obs_CHF),} \hlkwc{xlab} \hlstd{=} \hlstr{"Periods ahead"}\hlstd{,} \hlkwc{lty} \hlstd{=} \hlkwd{c}\hlstd{(}\hlkwd{rep}\hlstd{(}\hlnum{1}\hlstd{,} \hlnum{3}\hlstd{),} \hlnum{2}\hlstd{),}
\hlstd{+  }  \hlkwc{col} \hlstd{=} \hlkwd{c}\hlstd{(}\hlstr{"gray80"}\hlstd{,} \hlstr{"black"}\hlstd{,} \hlstr{"gray80"}\hlstd{,} \hlstr{"red"}\hlstd{))}
\end{alltt}
\end{kframe}
\end{knitrout}

\subsection{Rolling window estimation}

Inspired by \fct{ugarchroll} in the \proglang{R} package \pkg{rugarch}~\citep{rugarch}, we introduce the suite of wrapper functions \fct{svsample\_roll}, \fct{svtsample\_roll}, \fct{svlsample\_roll}, and \fct{svtlsample\_roll}, built around their corresponding routines \fct{svsample}, \fct{svtsample}, \fct{svlsample}, and, respectively, \fct{svtlsample}, to simplify rolling window estimation of SV models.
In this estimation method, either a fixed width time window is \emph{moving} through the time series or a sequence of \emph{expanding} time windows with the same starting time point covers larger and larger chunks of the observations, and the same model is estimated in all time windows independently.
Next, each estimated model is employed for out-of-sample prediction, typically one day to one week ahead of the time window.
Lastly, the set of predicted values might be used to evaluate the model fit.

In Bayesian statistics, a natural approach for assessing the predictive power of a model is through its \emph{posterior predictive distribution}.
Its density, also called the \emph{predictive density}, is defined as
\begin{equation}\label{eq:predictive-density}
  p(y_{t+1}\mid\bm{y^o}_{[1:t]})=\int_{\bm{\mathcal{K}}}p(y_{t+1}\mid\bm{y^o}_{[1:t]}, \bm\kappa) p(\bm\kappa\mid\bm{y^o}_{[1:t]}) \text{d}\bm\kappa,
\end{equation}
where $\bm\kappa$ collects all unobserved variables, i.e., $\bm\kappa=(\mupar,\phipar,\sigmapar,\rhopar,\nupar,\hvec,\bm\beta)^\top$ in the most general case of SVtl, and the domain of integration $\bm{\mathcal{K}}$ is the set of all possible values for $\bm\kappa$.
We follow~\citet{geweke2010comparing} in our notation by using a superscript $\bm o$ for the vector of observed values $\bm{y^o}_{[1:t]}=(y_1,y_2,\dots,y_t)^\top$.
Equation~\ref{eq:predictive-density} can be seen as the integration of the predictive likelihood over the posterior distribution of all parameters and therefore it accounts for posterior parameter uncertainty for the predicted values.

The integral in Equation~\ref{eq:predictive-density} has no closed form and its dimensionality increases with $t$; it is intractable.
Hence, we rely on Monte Carlo integration and we simulate from the posterior predictive distribution.
For the evaluation of the predictive density at an observation $x=y^o_{t+1}$, called the \emph{predictive likelihood}, we apply the computation
\begin{equation}\label{eq:predictive-simulation}
  p(x\mid\bm{y^o}_{[1:t]}) \approx \frac1{M}\sum_{m=1}^M p(x\mid\bm{y^o}_{[1:t]},\bm{\kappa^{(m)}}),
\end{equation}
where $\bm{\kappa^{(m)}}$ denotes the $m$th posterior sample from the estimation procedure of the SV model.
For other applications, the quantiles of the posterior predictive distribution, henceforth the \emph{predictive quantiles}, might be of interest.
We estimate the $q\%$ quantile through random variates simulated from $y_{t+1}\sim p(y_{t+1}\mid\bm{y^o}_{[1:t]})$, which we acquire by repeating two steps for $m=1,\dots,M$:
\begin{itemize}
  \item[Step 1.] Simulate $\bm{\kappa^{(m)}}$ from the SV posterior $p(\bm\kappa\mid\bm{y^o}_{[1:t]})$, and
  \item[Step 2.] Simulate $y^{(m)}_{t+1}$ from $p(y_{t+1}\mid\bm{y^o}_{[1:t]}, \bm\kappa^{(m)})$.
\end{itemize}
Lastly, we take the $q\%$ quantile of the sample vector $(y^{(1)}_{t+1}, y^{(2)}_{t+1}, \dots, y^{(M)}_{t+1})^\top$ as the approximate $q\%$ quantile of the predictive density.
We implement the estimation of both the predictive likelihood and predictive quantiles in \stochvol{}.

All four rolling window routines \fct{svsample\_roll}, \fct{svtsample\_roll}, \fct{svlsample\_roll}, and \fct{svtlsample\_roll} bear the same programming interface.
They expect as their first argument the input data $\yvec_{[1:L]}$, which is of length $L$.
For estimating the SV model in each time window $j=1,\dots,J$ in the moving or expanding window scheme, the sub-vector $\yvec_{[j:(t+j-1)]}$, or, respectively, $\yvec_{[1:(t+j-1)]}$, is taken as data and is used to predict $n_\text{ahead}\ge1$ time steps ahead.
The width $t$ of the first time window can be determined from $L$, $J$, and $n_\text{ahead}$.
The following example demonstrates how the rolling window sampling routines can be called in \stochvol{}.
\begin{knitrout}
\definecolor{shadecolor}{rgb}{0.969, 0.969, 0.969}\color{fgcolor}\begin{kframe}
\begin{alltt}
\hlstd{R> }\hlkwd{set.seed}\hlstd{(}\hlnum{5}\hlstd{)}
\hlstd{R> }\hlstd{res} \hlkwb{<-} \hlkwd{svsample_roll}\hlstd{(CHF_logret,} \hlkwc{n_ahead} \hlstd{=} \hlnum{1}\hlstd{,} \hlkwc{forecast_length} \hlstd{=} \hlnum{30}\hlstd{,}
\hlstd{+  }  \hlkwc{refit_window} \hlstd{=} \hlstr{"moving"}\hlstd{,} \hlkwc{calculate_quantile} \hlstd{=} \hlkwd{c}\hlstd{(}\hlnum{0.01}\hlstd{,} \hlnum{0.05}\hlstd{),}
\hlstd{+  }  \hlkwc{calculate_predictive_likelihood} \hlstd{=} \hlnum{TRUE}\hlstd{)}
\end{alltt}
\end{kframe}
\end{knitrout}
Argument \code{n\_ahead} is used to set $n_\text{ahead}$, \code{forecast\_length} is used to set $J$, and \code{refit\_window} expects either \code{"moving"} or \code{"expanding"} to set the rolling window scheme to moving or expanding, respectively.
Argument \code{calculate\_quantile} expects a vector of numbers between 0 and 1; the numbers are interpreted as the quantiles to be predicted.
Furthermore, if \code{calculate\_predictive\_likelihood} is set to \code{TRUE}, the function estimates the predictive likelihood.
Lastly, the output \code{res} is a list of length $J$, i.e., one element for each time window.
It contains the respective posterior quantile and predictive likelihood results together with all posterior parameter draws for $\bm\kappa$.

\subsection{Specifying the prior hyperparameters}
As discussed in Section~\ref{sec:svpriors}, the prior distributions need to be specified before the estimation process can start.
Concerning the common model parameters $\mupar$, $\phipar$, and $\sigmapar$, all of \fct{svsample}, \fct{svtsample}, \fct{svlsample}, and \fct{svtlsample} expect through their input arguments \code{priormu}, \code{priorphi}, and \code{priorsigma} values for $(\mupriorf,\sqrt{\mupriors})$, $(\phipriorf,\phipriors)$, and $\sigmaprior$, respectively.
Furthermore, all sampling functions accept the argument \code{priorbeta} to set an independent prior for the regression coefficients by providing $(b_\beta,s_\beta)$, where $b_\beta$ and $s_\beta$ are the common mean and, respectively, the common standard deviation. For a general multivariate normal distribution, the \fct{specify\_priors} interface exists, which we detail later in this Section.
The prior for $\nupar$ can be influenced in \fct{svtsample} and \fct{svtlsample} by passing $\nuprior$ as the argument \code{priornu}.
Finally, \fct{svlsample} and \fct{svtlsample} take the \code{numeric} sequence $(\rhopriorf,\rhopriors)$ through the input argument \code{priorrho}.

The code snippet below shows all the default values of the prior hyperparameters. 
\begin{knitrout}
\definecolor{shadecolor}{rgb}{0.969, 0.969, 0.969}\color{fgcolor}\begin{kframe}
\begin{alltt}
\hlstd{R> }\hlkwd{svsample}\hlstd{(CHF_logret,} \hlkwc{priormu} \hlstd{=} \hlkwd{c}\hlstd{(}\hlnum{0}\hlstd{,} \hlnum{100}\hlstd{),} \hlkwc{priorphi} \hlstd{=} \hlkwd{c}\hlstd{(}\hlnum{5}\hlstd{,} \hlnum{1.5}\hlstd{),}
\hlstd{+  }  \hlkwc{priorsigma} \hlstd{=} \hlnum{1}\hlstd{,} \hlkwc{priorbeta} \hlstd{=} \hlkwd{c}\hlstd{(}\hlnum{0}\hlstd{,} \hlnum{10000}\hlstd{))}
\hlstd{R> }\hlkwd{svtsample}\hlstd{(CHF_logret,} \hlkwc{priormu} \hlstd{=} \hlkwd{c}\hlstd{(}\hlnum{0}\hlstd{,} \hlnum{100}\hlstd{),} \hlkwc{priorphi} \hlstd{=} \hlkwd{c}\hlstd{(}\hlnum{5}\hlstd{,} \hlnum{1.5}\hlstd{),}
\hlstd{+  }  \hlkwc{priorsigma} \hlstd{=} \hlnum{1}\hlstd{,} \hlkwc{priorbeta} \hlstd{=} \hlkwd{c}\hlstd{(}\hlnum{0}\hlstd{,} \hlnum{10000}\hlstd{),} \hlkwc{priornu} \hlstd{=} \hlnum{0.1}\hlstd{)}
\hlstd{R> }\hlkwd{svlsample}\hlstd{(CHF_logret,} \hlkwc{priormu} \hlstd{=} \hlkwd{c}\hlstd{(}\hlnum{0}\hlstd{,} \hlnum{100}\hlstd{),} \hlkwc{priorphi} \hlstd{=} \hlkwd{c}\hlstd{(}\hlnum{5}\hlstd{,} \hlnum{1.5}\hlstd{),}
\hlstd{+  }  \hlkwc{priorsigma} \hlstd{=} \hlnum{1}\hlstd{,} \hlkwc{priorbeta} \hlstd{=} \hlkwd{c}\hlstd{(}\hlnum{0}\hlstd{,} \hlnum{10000}\hlstd{),} \hlkwc{priorrho} \hlstd{=} \hlkwd{c}\hlstd{(}\hlnum{4}\hlstd{,} \hlnum{4}\hlstd{))}
\hlstd{R> }\hlkwd{svtlsample}\hlstd{(CHF_logret,} \hlkwc{priormu} \hlstd{=} \hlkwd{c}\hlstd{(}\hlnum{0}\hlstd{,} \hlnum{100}\hlstd{),} \hlkwc{priorphi} \hlstd{=} \hlkwd{c}\hlstd{(}\hlnum{5}\hlstd{,} \hlnum{1.5}\hlstd{),}
\hlstd{+  }  \hlkwc{priorsigma} \hlstd{=} \hlnum{1}\hlstd{,} \hlkwc{priorbeta} \hlstd{=} \hlkwd{c}\hlstd{(}\hlnum{0}\hlstd{,} \hlnum{10000}\hlstd{),} \hlkwc{priornu} \hlstd{=} \hlnum{0.1}\hlstd{,}
\hlstd{+  }  \hlkwc{priorrho} \hlstd{=} \hlkwd{c}\hlstd{(}\hlnum{4}\hlstd{,} \hlnum{4}\hlstd{))}
\end{alltt}
\end{kframe}
\end{knitrout}

As an alternative to the concise interface above, a broader set of prior distributions can be specified via an object created by the \fct{specify\_priors} function.
The function has an input argument for each model parameter: \code{mu}, \code{phi}, \code{sigma2}, \code{nu}, \code{rho}, \code{beta}, and additionally one for the variance of $\hpar_0$ called \code{latent0\_variance}.
There is a list of accompanying functions that create distributions in \stochvol{}:
\fct{sv\_beta} has arguments \code{shape1} and \code{shape2} and it is accepted for \code{phi} and \code{rho};
\fct{sv\_constant} has argument \code{value} and it is accepted for \code{mu}, \code{phi}, \code{sigma2}, \code{nu}, \code{rho}, and \code{latent0\_variance};
\fct{sv\_normal} has arguments \code{mean} and \code{sd} and it is accepted for \code{mu} and \code{phi};
\fct{sv\_multinormal} has arguments \code{mean} and either \code{sd} and \code{dim} or \code{precision}, and it is accepted for \code{beta};
\fct{sv\_exponential} has argument \code{rate} and it accepted for \code{nu};
\fct{sv\_gamma} has arguments \code{shape} and \code{rate} and it is accepted for \code{sigma2};
\fct{sv\_inverse\_gamma} has arguments \code{shape} and \code{scale} and it is accepted for \code{sigma2};
and \fct{sv\_infinity} has no arguments and it is accepted for \code{nu} hence turning the Student's $t$~distribution into a normal distribution.
Additionally, \code{latent0\_variance} accepts the character value \code{"stationary"}.
All four sampling methods accept the prior specification object through the input argument \code{priorspec}.

All input arguments for \code{specify\_priors} are optional, their default values and how they are used is seen below.
\begin{knitrout}
\definecolor{shadecolor}{rgb}{0.969, 0.969, 0.969}\color{fgcolor}\begin{kframe}
\begin{alltt}
\hlstd{R> }\hlstd{ps} \hlkwb{<-} \hlkwd{specify_priors}\hlstd{(}\hlkwc{mu} \hlstd{=} \hlkwd{sv_normal}\hlstd{(}\hlkwc{mean} \hlstd{=} \hlnum{0}\hlstd{,} \hlkwc{sd} \hlstd{=} \hlnum{100}\hlstd{),}
\hlstd{+  }  \hlkwc{phi} \hlstd{=} \hlkwd{sv_beta}\hlstd{(}\hlkwc{shape1} \hlstd{=} \hlnum{5}\hlstd{,} \hlkwc{shape2} \hlstd{=} \hlnum{1.5}\hlstd{),} \hlkwc{rho} \hlstd{=} \hlkwd{sv_constant}\hlstd{(}\hlnum{0}\hlstd{),}
\hlstd{+  }  \hlkwc{sigma2} \hlstd{=} \hlkwd{sv_gamma}\hlstd{(}\hlkwc{shape} \hlstd{=} \hlnum{0.5}\hlstd{,} \hlkwc{rate} \hlstd{=} \hlnum{0.5}\hlstd{),} \hlkwc{nu} \hlstd{=} \hlkwd{sv_infinity}\hlstd{(),}
\hlstd{+  }  \hlkwc{beta} \hlstd{=} \hlkwd{sv_multinormal}\hlstd{(}\hlkwc{mean} \hlstd{=} \hlnum{0}\hlstd{,} \hlkwc{sd} \hlstd{=} \hlnum{10000}\hlstd{,} \hlkwc{dim} \hlstd{=} \hlnum{1}\hlstd{),}
\hlstd{+  }  \hlkwc{latent0_variance} \hlstd{=} \hlstr{"stationary"}\hlstd{)}
\hlstd{R> }\hlkwd{svsample}\hlstd{(CHF_logret,} \hlkwc{priorspec} \hlstd{= ps)}
\end{alltt}
\end{kframe}
\end{knitrout}

\subsection{Setting up the Markov chain}

When conducting Bayesian inference using an MCMC sampling scheme, the number of draws from the posterior distribution, the length of the so-called burn-in phase, the initial values of the Markov chain, and the various strategies of storing the results are all of general interest.
The input arguments \code{draws} and \code{burnin} settle the first two points.
A sample size of \code{burnin + draws} is acquired from the posterior distribution out of which the first \code{burnin} number of draws are thrown away.
The default is to draw 10000 elements after a burnin of 1000 for SV models without leverage, and draw 20000 elements after a burnin of 2000 for SV models with leverage, which in our experience is enough for most applications.

As for the initial values, \code{startpara} and \code{startlatent} provide a way to set them.
The argument \code{startpara} is expected to be a named \code{list} mapping parameter names to starting values, and \code{startlatent} must be a sequence of length $\dimy$ that contains starting values for $\hvec$.
Default values are set to be the prior mean for $\phipar$, $\sigmapar$, $\nupar$, and $\rhopar$, these have only minor influence on the Markov chain.
The default value for $\bm\beta$ is the ordinary least squares estimator $\bm{(X^\top X)^{-1}\bm{X^\top}}\yvec$, where $\bm X$ denotes the regression design matrix and $\yvec$ denotes the vector of observations.
After setting $\bm\beta$, the level of log-variance $\mupar$ is initialized according to the Bayesian linear regression
\begin{equation}\label{eq:initmu}
  \begin{split}
    \log(y_t^2) &= \mupar+\xi_t, \\
    \mupar &\sim \Normal{\mupriorf}{\mupriors},
  \end{split}
\end{equation}
where $\xi_t\sim\Normal{-1.27}{4.934}$.
Equation~\ref{eq:initmu} results from the first line of Equation~\ref{eq:vanillasv} by fixing $h_t$ at its stationary expected value $\mupar$ and then taking $x\mapsto\log(x^2)$ of both sides.
The homoskedastic error term $\xi_t$ is acquired as the Laplace approximation to $\log(\varepsilon_t^2)$~\citep{harvey1996estimation}.
At the end, by default all values of the vector \code{startlatent} are set to the initial value of $\mupar$.

It is customary to start independent Markov chains in parallel and \stochvol{} provides facilities for that in all of its sampling procedures.
The argument \code{n\_chains} is expected to be a positive integer, it sets the number of independent chains.
Additionally, arguments \code{parallel}, \code{n\_cpus}, and \code{cl} can be used to control parallelism used by \stochvol{}.
To overwrite the default sequential execution strategy, \code{parallel} is to be set either to \code{"snow"}, to employ the so-called ``SNOW'' clusters, or to \code{"multicore"} to use the ``multicore'' type computation~\citep{rlanguage}.
Next, argument \code{n\_cpus} should be set to the physical number of parallel processing units to be used.
Finally, in case ``SNOW'' is applied, the sampling routines optionally accept an already running ``SNOW'' cluster through argument \code{cl}.

As mentioned earlier, the sampling algorithms for the latent states $\hvec$ in \stochvol{} rely on a Gaussian mixture approximation as in~\citet{omori2007stochastic} and~\citet{kastner2014ancillarity}.
The approximation tends to be very good, therefore the default setting is not to correct for model misspecification.
However, this correction can be enabled in all of the sampling routines through the \code{expert} argument as shown for \fct{svsample} in the following.

Lastly, \stochvol{} provides three ways to economize storage during and after the execution of the sampler.
Setting the \code{integer} argument \code{thinpara} to $\iota$ tells the sampler to store only every $\iota$th draw of the vector of parameters, and supplying a value for \code{thinlatent} does the same for $\hvec$.
Finally, one has the opportunity not to store the full vector $\hvec$ but only its last value by setting \code{keeptime = "last"}.
The default behavior is to store every draw after the burn-in phase.

\section[The factorstochvol package]{The \factorstochvol{} package} \label{sec:factorstochvol}

The most common workflow of using \factorstochvol{} for fitting multivariate factor SV models consists of the following steps:
(1) Prepare the data,
(2) decide on an identification structure,
(3) specify the prior hyperparameters,
(4) run the sampler,
(5) investigate the output and visualize the results, and
(6) predict (if required).
These steps are described in detail in the following sections.

\subsection{Preparing the data} \label{sec:fsvuse1}

\begin{figure}[t]
  \centering
  \includegraphics[trim=5 40 20 40, clip, width=\textwidth]{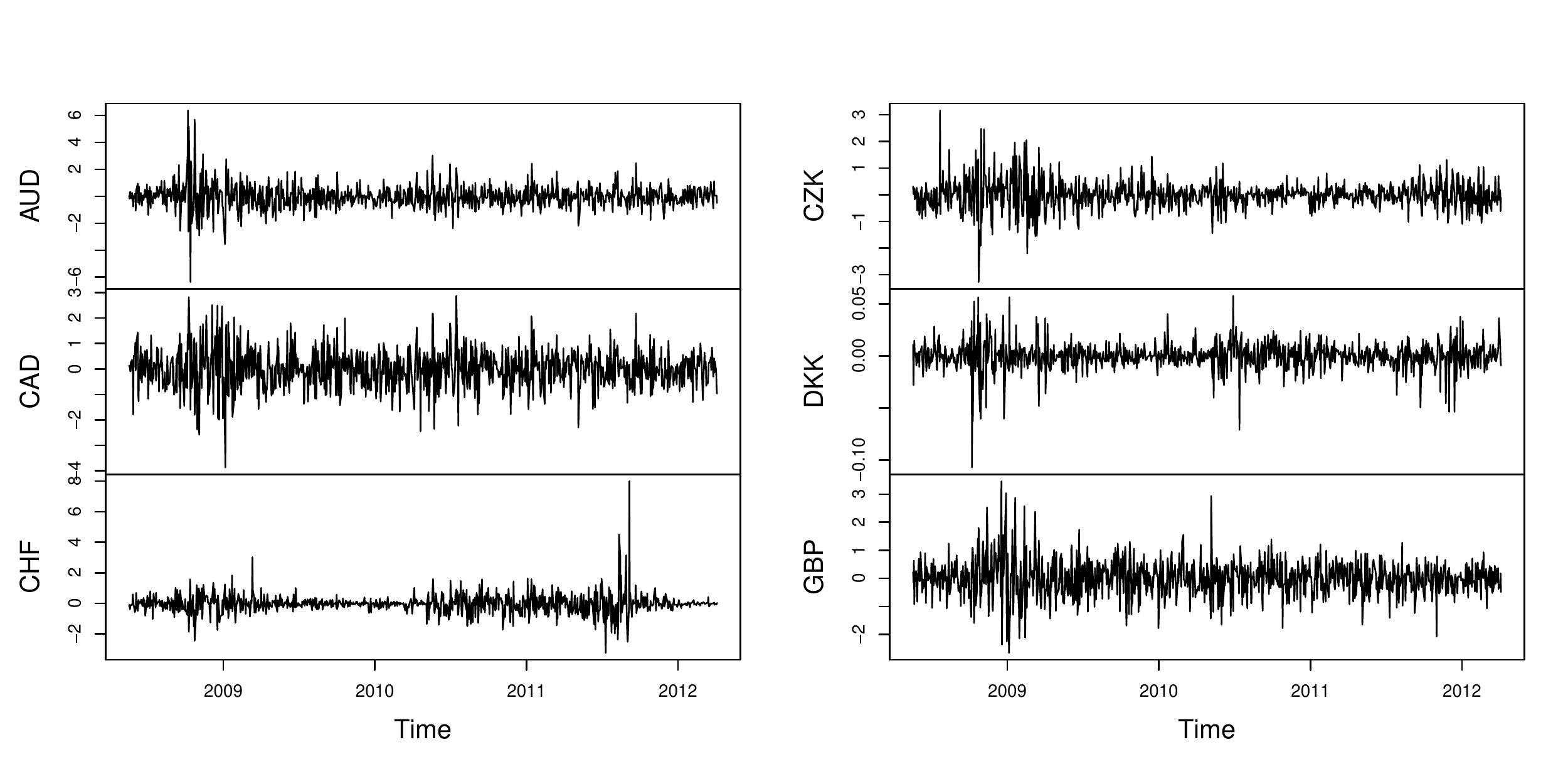}
  \caption{Percentage log returns of six EUR exchange rates.}
  \label{fig:exrates}
\end{figure}

The workhorse in \factorstochvol{} is the sampling function \fct{fsvsample}. It expects the data to come in form of a matrix $\bm Y=(\yvec_1,\dots,\yvec_\leny)^\top$ with $\leny$ rows and $\dimy$ columns.
For illustration, we use the exchange rate data set in \stochvol{} which contains $3140$ daily observations of exchanges rates for $23$ currency pairs against EUR, ranging from March 3, 2000 to April 4, 2012.
To keep the analysis simple and computation times moderate, we however only model the last $1001$ days of the first six series in alphabetical order (Australian dollar, Canadian dollar, Swiss franc, Czech koruna, Danish krone, Great British pound) for further analysis.
Instead of using the nominal exchange rates we compute log returns. This leaves us with a data set of size $n = 1000$ and $m = 6$.
The data is prepared using the code snippet below and visualized in Figure~\ref{fig:exrates} using the \pkg{zoo} package \citep{zoo}.
\begin{knitrout}
\definecolor{shadecolor}{rgb}{0.969, 0.969, 0.969}\color{fgcolor}\begin{kframe}
\begin{alltt}
\hlstd{R> }\hlkwd{library}\hlstd{(}\hlstr{"factorstochvol"}\hlstd{)}
\hlstd{R> }\hlkwd{library}\hlstd{(}\hlstr{"zoo"}\hlstd{)}
\hlstd{R> }\hlkwd{data}\hlstd{(}\hlstr{"exrates"}\hlstd{,} \hlkwc{package} \hlstd{=} \hlstr{"stochvol"}\hlstd{)}
\hlstd{R> }\hlstd{m} \hlkwb{<-} \hlnum{6}
\hlstd{R> }\hlstd{n} \hlkwb{<-} \hlnum{1000}
\hlstd{R> }\hlstd{y} \hlkwb{<-} \hlnum{100} \hlopt{*} \hlkwd{logret}\hlstd{(}\hlkwd{tail}\hlstd{(exrates[,} \hlkwd{seq_len}\hlstd{(m)], n} \hlopt{+} \hlnum{1}\hlstd{))}
\hlstd{R> }\hlstd{y} \hlkwb{<-} \hlkwd{zoo}\hlstd{(y,} \hlkwc{order.by} \hlstd{=} \hlkwd{tail}\hlstd{(exrates}\hlopt{$}\hlstd{date, n))}
\hlstd{R> }\hlkwd{plot}\hlstd{(y,} \hlkwc{main} \hlstd{=} \hlstr{""}\hlstd{,} \hlkwc{xlab} \hlstd{=} \hlstr{"Time"}\hlstd{)}
\end{alltt}
\end{kframe}
\end{knitrout}

\subsection{Deciding on an identification structure}\label{sec:fsvuse2}

The likelihood in factor models is invariant to certain factor transformations such as reordering of factors and their loadings or sign switches thereof. In addition to this, it is often multimodal. Consequently, identifying the factor loadings is far from  trivial.
The most common way to address this issue in factor SV models is to impose a lower-diagonal factor loadings matrix where all elements above the diagonal are set to zero \citep[e.g.,][]{aguilar2000bayesian, chib2006analysis, han2006asset, zhou2014bayesian}.
To use this constraint in \factorstochvol, the argument \code{restrict = "upper"} can be passed to the main sampling function \fct{fsvsample}.
Evidently, this practice imposes an order dependence, as, e.g., the first variable is not allowed to load on anything else but the first factor.

A rather ad hoc method for automatically ordering the data is implemented in the helper function \fct{preorder}.
After a maximum likelihood factor model fit to the data (using \fct{factanal} from the \pkg{stats} package with the default \code{varimax} rotation), the series are ordered as follows: The variable with the highest loading on factor 1 is placed first, the variable with the highest loading on factor 2 second (unless this variable is already placed first, in which case the variable with the second highest loading is taken), et cetera.
For the data set at hand, this would imply the following ordering for a two-factor model.
\begin{knitrout}
\definecolor{shadecolor}{rgb}{0.969, 0.969, 0.969}\color{fgcolor}\begin{kframe}
\begin{alltt}
\hlstd{R> }\hlkwd{preorder}\hlstd{(y,} \hlkwc{factors} \hlstd{=} \hlnum{2}\hlstd{)}
\end{alltt}
\begin{verbatim}
## [1] 2 3 1 4 5 6
\end{verbatim}
\end{kframe}
\end{knitrout}
According to this algorithm, the second series should be placed first and the third series should be placed second. Thereafter, the alphabetical ordering remains.

To achieve this effect without reordering the data, a \code{logical} matrix of size $\dimy \times r$ can be passed to \fct{fsvsample} via \code{restrict}, where the entry \code{TRUE} means that this element is restricted to zero; \code{FALSE} means that it is to be estimated from the data.
Similarly to \fct{preorder}, the function \fct{findrestrict} tries to automate this procedure.
Again, the maximum likelihood estimates from a static factor analysis are used; however, \fct{findrestrict} uses a slightly different algorithm than the one above:
The variable with the lowest absolute loadings on factors $2, 3, \dots, r$ (relative to factor 1) is determined to lead the first factor, the variable with the lowest absolute loadings on factors $3, 4, \dots, r$ (relative to factors 1 and 2) is placed second, et cetera.
Below is the result for the data set at hand.
\begin{knitrout}
\definecolor{shadecolor}{rgb}{0.969, 0.969, 0.969}\color{fgcolor}\begin{kframe}
\begin{alltt}
\hlstd{R> }\hlkwd{findrestrict}\hlstd{(y,} \hlkwc{factors} \hlstd{=} \hlnum{2}\hlstd{)}
\end{alltt}
\begin{verbatim}
##       [,1]  [,2]
## [1,] FALSE FALSE
## [2,] FALSE  TRUE
## [3,] FALSE FALSE
## [4,] FALSE FALSE
## [5,] FALSE FALSE
## [6,] FALSE FALSE
\end{verbatim}
\end{kframe}
\end{knitrout}
If \fct{fsvsample} is called with the argument \code{restrict = "auto"}, it automatically invokes \fct{findrestrict} with the appropriate number of factors.
Using \code{restrict = "none"} (the default) causes the sampler not to place any constraints on the loadings matrix; thus, the resulting posterior draws may be unstable or suffer from multiple local modes.
If, however, inference on the factor loadings themselves is not the primary concern of the analysis, leaving the factor loadings unidentified may be the preferred option.
This is in particular the case when inference for the covariance matrix is sought, as this only depends on $\Loadings$ through the rotation-invariant transformation of Equation~\ref{eq:vardecomp}.
For a more elaborate discussion of these issues, we refer the reader to \citet{sentana2001identification} who discuss automatic identification through hetero\-skedasticity. A comparison of log predictive scores under different identification schemes for factor SV models is given in \citet{kastner2017efficient}; see also \citet{fruhwirth2018sparse} for related issues in static factor models.
To continue with the current example, we chose not to place any a priori restrictions on the factor loadings matrix while using a row-wise normal-gamma shrinkage prior on the factor loadings matrix \citep[cf.][]{kastner2019sparse}.

\subsection{Specifying prior hyperparameters} \label{sec:fsvuse3}

Apart from the obvious prior choice about the number of factors and the identification scheme discussed above, a number of hyperparameter choices are available in \factorstochvol.
Regarding the log-variance processes, the interface is analogous to that of \fct{svsample}.
In the following, $i=1,\dots,\dimy$ and $j=1,\dots,\nfac$ index the idiosyncratic and the factor log-variance processes, respectively.
The pair of common prior hyperparameters $(b_\beta, B_\beta)$ can be passed as a sequence of length two to \code{priorbeta}.
The common prior of $\muidi_i$ can be set by passing a sequence of length two -- the mean and the standard deviation of the normal distribution -- to \code{priormu};
the common priors of $\phiidi_i$ and $\phifac_j$ can also be set by passing sequences of length two -- the parameters of the corresponding beta distribution -- to \code{priorphiidi} and to \code{priorphifac}, respectively;
similarly, the common priors of $\sigmaidi_i$ and $\sigmafac_j$ can be specified via the arguments \code{priorsigmaidi} and \code{priorsigmafac}, respectively, that accept as positive numbers the scale $\sigmaprior$ of the corresponding gamma distribution.

As discussed in Section~\ref{sec:fsvpriors}, \factorstochvol{} offers three specifications as priors for $\Loadings$, controlled through the argument \code{priorfacloadtype}.
To use the first option (\code{priorfacloadtype = "normal"}), one needs to fix the values of $\tau_{ij}$ a priori.
The user can pass these fixed values to \fct{fsvsample} via the argument \code{priorfacload}, either as an $\dimy\times\nfac$ matrix with positive entries or as a single positive number which will be recycled accordingly.
For the second option, the normal gamma prior with row-wise or column-wise shrinkage (\code{priorfacloadtype = "rowwiseng"} and \code{priorfacloadtype = "colwiseng"}, respectively), the value of argument \code{priorfacload} is then interpreted as the shrinkage parameter $a$.
Both specifications of the normal gamma prior need the values $c$ and $d$.
They can be set as a two-element vector passed to the argument \code{priorng}.

\subsection{Running the MCMC sampler}

Running the sampler corresponds to invoking \fct{fsvsample}. Apart from the prior settings discussed above, its most important arguments are listed below with the default value in brackets. For a complete list of all arguments and more details, see \code{?fsvsample}.
\begin{itemize}
  \item \code{y}: the data;
  \item \code{factors} [\code{1}]: the number of factors;
  \item \code{draws} [\code{1000}]: the number of MCMC samples to be drawn after burnin;
  \item \code{thin} [\code{1}]: the amount of thinning (every \code{thin}\textsuperscript{th} draw is kept);
  \item \code{burnin} [\code{1000}]: the length of the burn-in period, i.e., the number of MCMC draws to be discarded before the samples are considered to emerge from the stationary distribution,
  \item \code{zeromean} [\code{TRUE}]: a logical value indicating whether $\bm\beta$ is to be estimated from the data or whether $\bm\beta$ is set to zero (the default);
  \item \code{keeptime} [\code{"last"}]: either \code{"all"}, meaning that all latent log volatilities are being monitored at all points in time, or \code{"last"}, meaning that the latent log volatility draws are only stored at $t = n$, the last point in time; the latter setting is the default to avoid excessive memory usage in higher dimensions;
  \item \code{heteroskedastic} [\code{TRUE}]: indicator(s) to turn off stochastic volatility for the idiosyncratic variances, the factor variances, or both;
  \item \code{samplefac} [\code{TRUE}]: indicator to turn off sampling of the factors; useful to work with observed instead of latent factors \citep[see][for a use case of this]{kastner2019sparse};
  \item \code{runningstore} [\code{6}]: to avoid having to store all MCMC draws, \code{fsvsample}'s default is to compute and store the first two ergodic moments of some interesting quantities (namely log variances, factors, volatilities, covariance matrices, correlation matrices, communalities) only; the default (\code{runningstore = 6}) is to compute and store everything; however, one can set \code{runningstore} to a lower number to save computation time; the argument \code{runningstoremoments} [\code{2}] can further be used to modify the number of moments to be stored;
  \item \code{runningstorethin} [\code{10}]: indicates how often ergodic moments should be calculated, where \code{1} means that this should be done at every iteration and higher numbers lessen both runtime as well as accuracy;
  \item \code{quiet} [\code{FALSE}]: a logical indicator determining the verbosity of \code{fsvsample}.
\end{itemize}

For our illustrative example, most settings are left at their default values. The number of factors is increased from one to two, instead of 1000 we sample 10000 draws, we estimate a constant mean, a thinning of 10 is used, and \code{quiet} is set to \code{TRUE}.
\begin{knitrout}
\definecolor{shadecolor}{rgb}{0.969, 0.969, 0.969}\color{fgcolor}\begin{kframe}
\begin{alltt}
\hlstd{R> }\hlkwd{set.seed}\hlstd{(}\hlnum{1}\hlstd{)}
\hlstd{R> }\hlstd{res} \hlkwb{<-} \hlkwd{fsvsample}\hlstd{(y,} \hlkwc{factors} \hlstd{=} \hlnum{2}\hlstd{,} \hlkwc{draws} \hlstd{=} \hlnum{10000}\hlstd{,} \hlkwc{zeromean} \hlstd{=} \hlnum{FALSE}\hlstd{,}
\hlstd{+  }  \hlkwc{thin} \hlstd{=} \hlnum{10}\hlstd{,} \hlkwc{quiet} \hlstd{=} \hlnum{TRUE}\hlstd{)}
\end{alltt}
\end{kframe}
\end{knitrout}

\subsection{Investigating the output and visualizing the results}

The resulting object 
\begin{knitrout}
\definecolor{shadecolor}{rgb}{0.969, 0.969, 0.969}\color{fgcolor}\begin{kframe}
\begin{alltt}
\hlstd{R> }\hlstd{res}
\end{alltt}
\begin{verbatim}
## 
## Fitted factor stochastic volatility object with
##   -       6 series
##   -       2 factor(s)
##   -    1000 timepoints
##   -   10000 MCMC draws
##   -      10 thinning
##   -    1000 burn-in
\end{verbatim}
\end{kframe}
\end{knitrout}
holds a rich amount of information. In particular, it contains
\begin{itemize}
  \item draws of certain posterior quantities such as the factors $\fvec$, the factor loadings $\Loadings$, the various factor and idiosyncratic SV parameters, the latent factor and idiosyncratic log variances $\bm{\hfac}$ and $\bm{\hidi}$, and the intercept $\bm\beta$,
  \item configuration settings such as the number of draws, potential restrictions on the loadings matrix, prior hyperparameters, etc.,
  \item running moments (such as means and standard deviations) of quantities of interest, depending on the values of \code{runningstore} and \code{runningstoremoments} specified when calling \fct{fsvsample},
  \item the data input $\bm{y}$.
\end{itemize}
For more details, please investigate \code{str(res)} and/or \code{?fsvsample}.

\begin{figure}[tp]
  \centering
  \includegraphics[width=\textwidth, clip, trim=5 5 10 10]{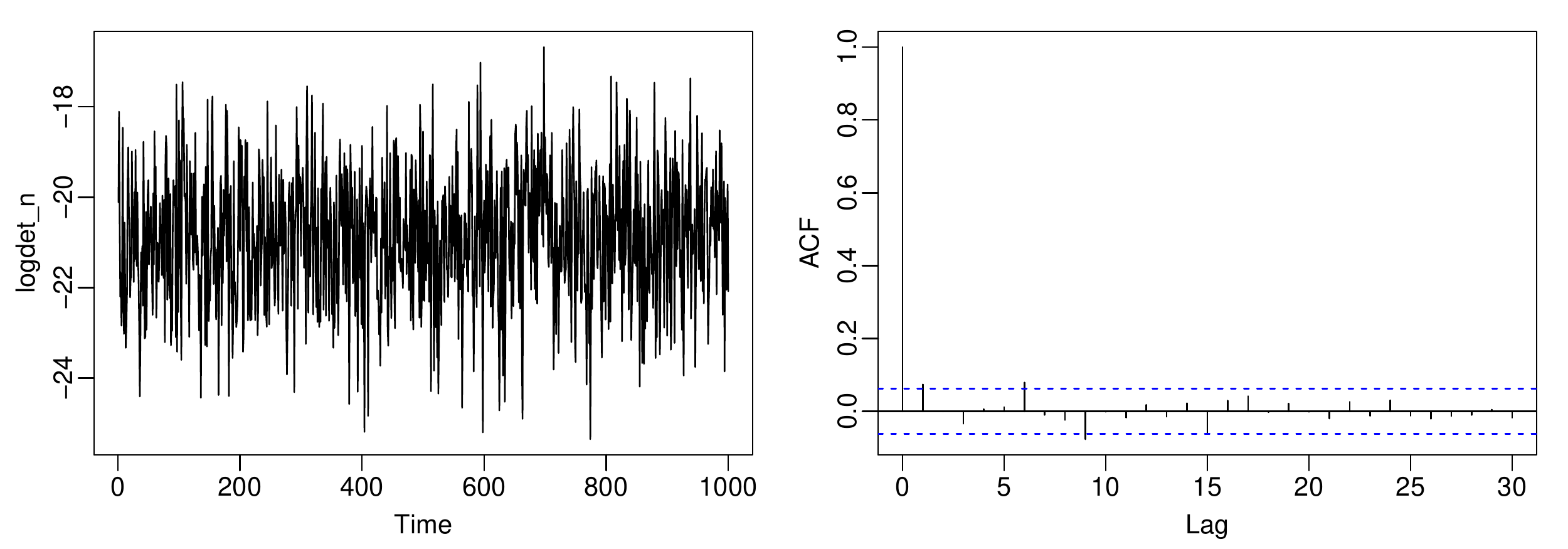}
  \caption{Trace plot and empirical autocorrelation function of the log determinant of the model-implied covariance matrix at $t = n$.}
  \label{fig:logdetcovn}
\end{figure}

Using \fct{covmat}, one can extract the MCMC draws of the implied covariance matrices for all points in time which have been stored during sampling. By default, this is the last point in time (\code{keeptime = "last"}), and thus
\begin{knitrout}
\definecolor{shadecolor}{rgb}{0.969, 0.969, 0.969}\color{fgcolor}\begin{kframe}
\begin{alltt}
\hlstd{R> }\hlkwd{dim}\hlstd{(cov_n} \hlkwb{<-} \hlkwd{covmat}\hlstd{(res))}
\end{alltt}
\begin{verbatim}
## [1]    6    6 1000    1
\end{verbatim}
\end{kframe}
\end{knitrout}
shows that we have stored $1000$ posterior draws of a $6\times6$ covariance matrix at one point in time, $t = n = 1000$. To check convergence, one can take a look at the trace plot and the autocorrelation function of the log determinant, i.e.,
\begin{knitrout}
\definecolor{shadecolor}{rgb}{0.969, 0.969, 0.969}\color{fgcolor}\begin{kframe}
\begin{alltt}
\hlstd{R> }\hlstd{logdet} \hlkwb{<-} \hlkwa{function} \hlstd{(}\hlkwc{x}\hlstd{)} \hlkwd{log}\hlstd{(}\hlkwd{det}\hlstd{(x))}
\hlstd{R> }\hlstd{logdet_n} \hlkwb{<-} \hlkwd{apply}\hlstd{(cov_n[,,,}\hlnum{1}\hlstd{],} \hlnum{3}\hlstd{, logdet)}
\hlstd{R> }\hlkwd{ts.plot}\hlstd{(logdet_n)}
\hlstd{R> }\hlkwd{acf}\hlstd{(logdet_n,} \hlkwc{main} \hlstd{=} \hlstr{""}\hlstd{)}
\end{alltt}
\end{kframe}
\end{knitrout}
The results are visualized in Figure~\ref{fig:logdetcovn}; decent mixing for this quantity is apparent.

To assess the mixing speed of each individual covariance matrix element, one can check, e.g., the estimated effective sample size (out of $1000$ draws kept) which is implemented in \pkg{coda}. Again, no major convergence problems are apparent.
\begin{knitrout}
\definecolor{shadecolor}{rgb}{0.969, 0.969, 0.969}\color{fgcolor}\begin{kframe}
\begin{alltt}
\hlstd{R> }\hlkwd{round}\hlstd{(}\hlkwd{apply}\hlstd{(cov_n,} \hlnum{1}\hlopt{:}\hlnum{2}\hlstd{, coda}\hlopt{::}\hlstd{effectiveSize))}
\end{alltt}
\begin{verbatim}
##      [,1] [,2] [,3] [,4] [,5] [,6]
## [1,]  779 1000  735  721  776 1000
## [2,] 1000  893  575  661  786 1000
## [3,]  735  575  700  597  454  415
## [4,]  721  661  597 1127  527  450
## [5,]  776  786  454  527 1000  764
## [6,] 1000 1000  415  450  764  859
\end{verbatim}
\end{kframe}
\end{knitrout}

Assuming that \code{runningstore} was set sufficiently high when sampling, several convenience functions can be used for quick visualizations without having to post-process the MCMC draws. For example, to visualize the time-varying correlation matrices, consider
\begin{knitrout}
\definecolor{shadecolor}{rgb}{0.969, 0.969, 0.969}\color{fgcolor}\begin{kframe}
\begin{alltt}
\hlstd{R> }\hlkwd{corimageplot}\hlstd{(res,} \hlkwc{these} \hlstd{=} \hlkwd{seq}\hlstd{(}\hlnum{1}\hlstd{, n,} \hlkwc{length.out} \hlstd{=} \hlnum{3}\hlstd{),} \hlkwc{plotCI} \hlstd{=} \hlstr{"circle"}\hlstd{,}
\hlstd{+  }  \hlkwc{plotdatedist} \hlstd{=} \hlnum{2}\hlstd{,} \hlkwc{date.cex} \hlstd{=} \hlnum{1.1}\hlstd{)}
\end{alltt}
\end{kframe}
\end{knitrout}
\begin{figure}[tp]
  \centering
  \includegraphics[width=\textwidth, clip, trim=22 185 0 174]{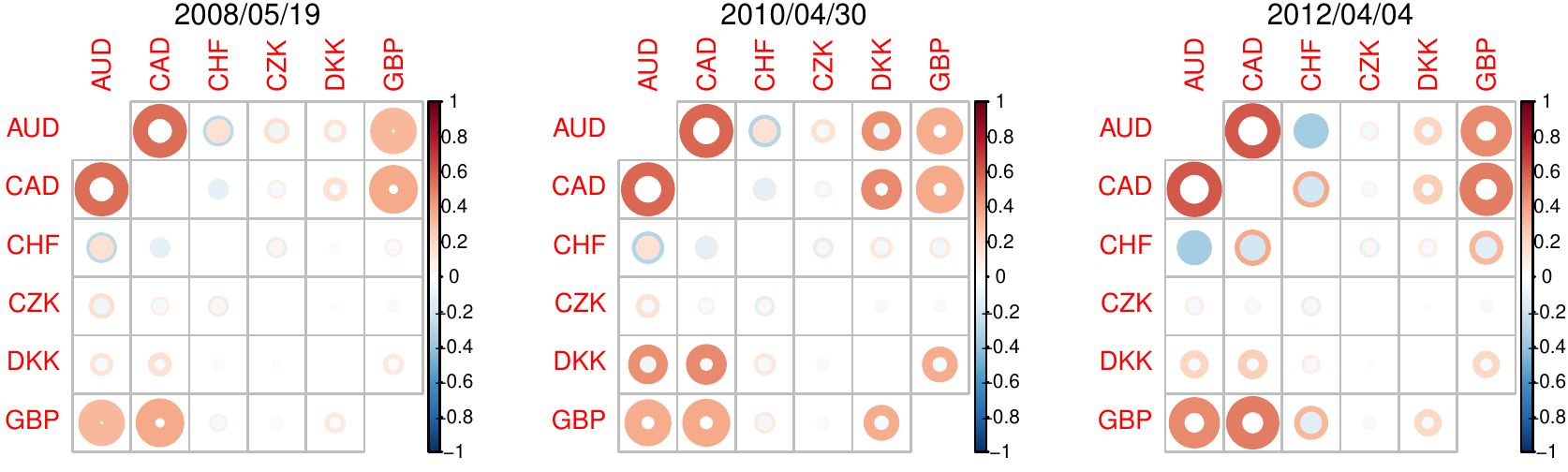}
  \caption{Three estimated correlation matrices and their posterior uncertainty depicted using circles.
    Inner (outer) radii of the circles illustrate to the posterior mean minus (plus) two standard deviations.
    Colors blue and red represent negative and positive values, respectively.
    Furthermore, the transparency of the circles represents the posterior mean.
  Note that the diagonal is left white as it trivially contains ones (without uncertainty).}
  \label{fig:corimageplot}
\end{figure}
which produces the three estimated posterior correlation matrices depicted in Figure~\ref{fig:corimageplot}. Setting \code{plotCI = "circle"} visualizes posterior uncertainty -- inner and outer radii correspond to the posterior mean plus/minus two standard deviations, respectively.

To get an idea about how the marginal volatilities evolve over time, \fct{voltimeplot} can be used. To exemplify,
\begin{knitrout}
\definecolor{shadecolor}{rgb}{0.969, 0.969, 0.969}\color{fgcolor}\begin{kframe}
\begin{alltt}
\hlstd{R> }\hlkwd{palette}\hlstd{(RColorBrewer}\hlopt{::}\hlkwd{brewer.pal}\hlstd{(}\hlnum{7}\hlstd{,} \hlstr{"Dark2"}\hlstd{)[}\hlopt{-}\hlnum{5}\hlstd{])}
\hlstd{R> }\hlkwd{voltimeplot}\hlstd{(res,} \hlkwc{legend} \hlstd{=} \hlstr{"top"}\hlstd{)}
\end{alltt}
\end{kframe}
\end{knitrout}
\begin{figure}[tp]
  \centering
  \includegraphics[width=\textwidth, clip, trim=0 10 2 10]{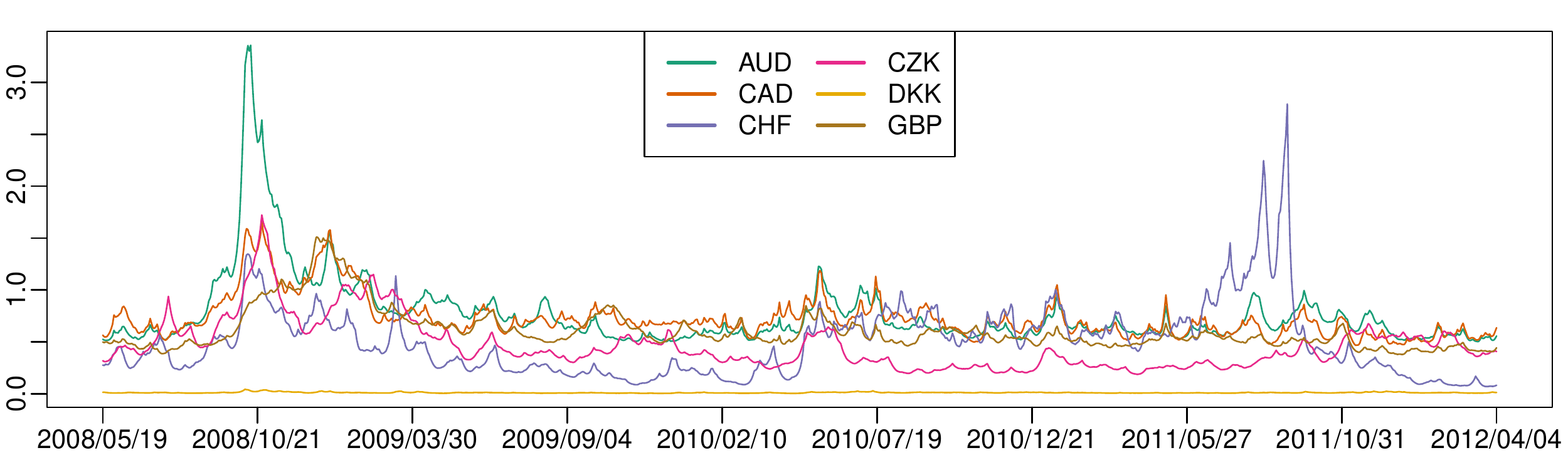}
  \caption{Posterior means of daily marginal volatilities in percent.}
  \label{fig:voltimeplot}
\end{figure}
yields the estimated volatilities in Figure~\ref{fig:voltimeplot}. The financial crisis of 2008 and the capping of CHF's appreciation in September 2011 are clearly visible, while DKK's volatility (relative to EUR) is practically zero. Note that \fct{voltimeplot} respects palette changes. In the above example, \pkg{RcolorBrewer} \citep{rcolorbrewer} is used. Moreover,
\begin{knitrout}
\definecolor{shadecolor}{rgb}{0.969, 0.969, 0.969}\color{fgcolor}\begin{kframe}
\begin{alltt}
\hlstd{R> }\hlkwd{palette}\hlstd{(RColorBrewer}\hlopt{::}\hlkwd{brewer.pal}\hlstd{(}\hlnum{6}\hlstd{,} \hlstr{"Dark2"}\hlstd{))}
\hlstd{R> }\hlkwd{cortimeplot}\hlstd{(res,} \hlnum{1}\hlstd{)}
\hlstd{R> }\hlkwd{cortimeplot}\hlstd{(res,} \hlnum{2}\hlstd{)}
\end{alltt}
\end{kframe}
\end{knitrout}
\begin{figure}[tp]
  \centering
  \includegraphics[width=\textwidth, clip, trim=4 4 0 4]{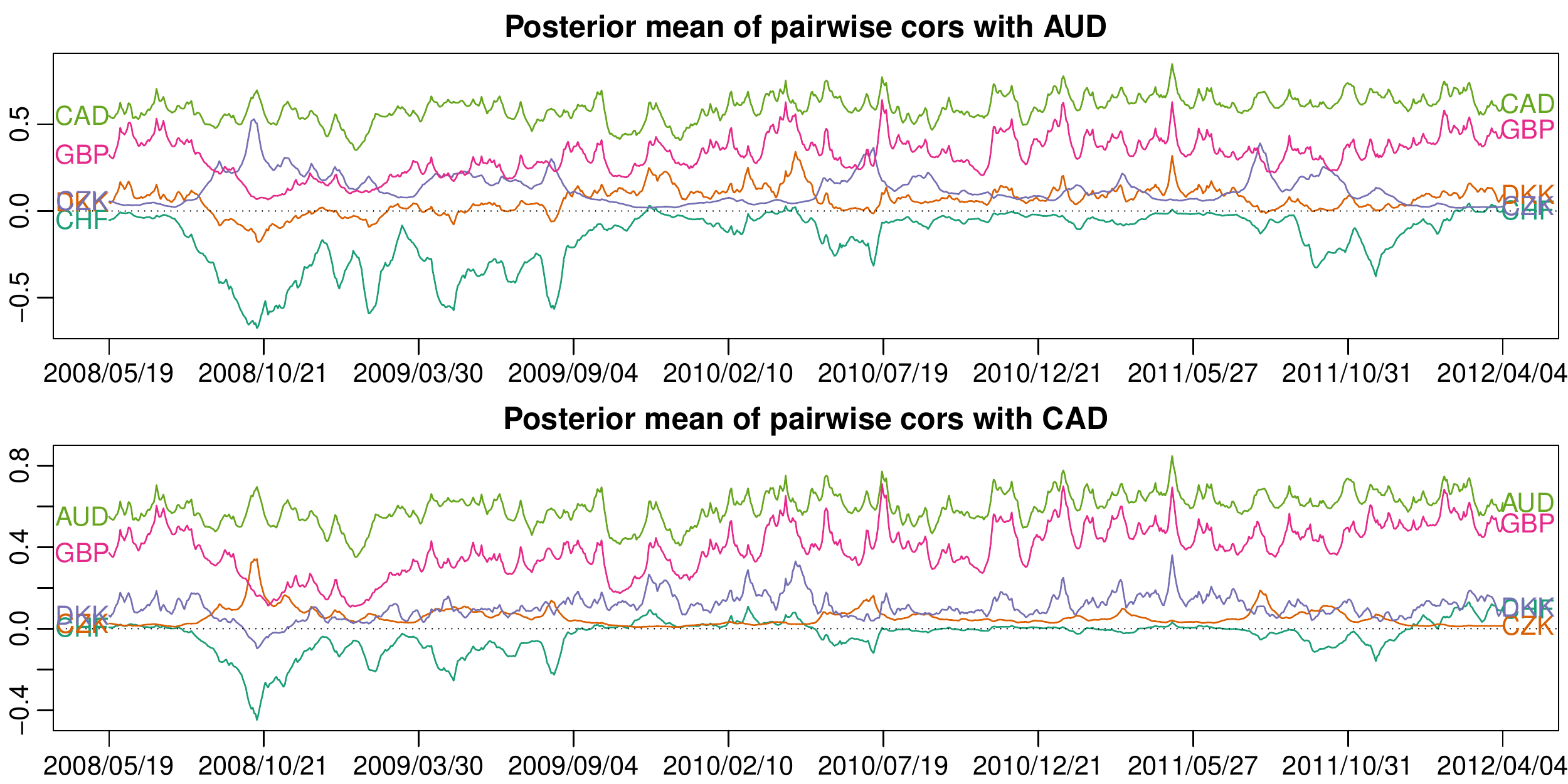}
  \caption{Posterior means of correlations with AUD (top panel) and CAD (bottom panel).}
  \label{fig:cortimeplot}
\end{figure}
yields the estimated pairwise correlations in Figure~\ref{fig:cortimeplot}. While, relative to EUR, the estimated correlation between AUD and CAD appears to be relatively stable over time, correlations with CHF can become negative at times. To visualize the \emph{communalities}, i.e., the proportions of variances explained through the latent factors, invoke
\begin{knitrout}
\definecolor{shadecolor}{rgb}{0.969, 0.969, 0.969}\color{fgcolor}\begin{kframe}
\begin{alltt}
\hlstd{R> }\hlkwd{comtimeplot}\hlstd{(res,} \hlkwc{maxrows} \hlstd{=} \hlnum{6}\hlstd{)}
\end{alltt}
\end{kframe}
\end{knitrout}
\begin{figure}[tp]
  \centering
  \includegraphics[width=\textwidth]{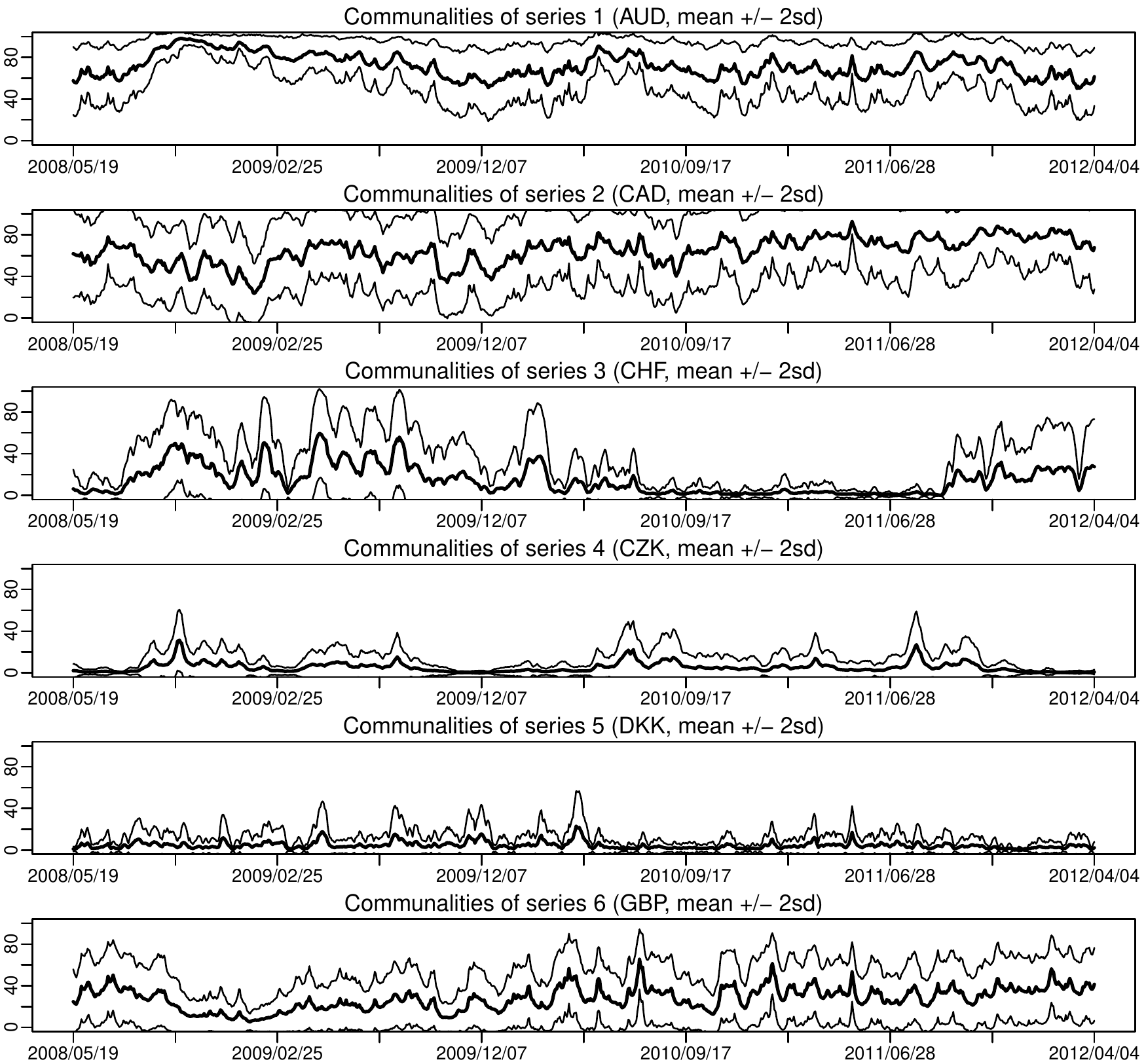}
  \caption{Communalities: Posterior means plus/minus two posterior standard deviations.
    The six panels correspond to the six observation series and they depict the percentage of volatility explained by the latent factors.
    The idiosyncratic volatility attributes for the unexplained part.}
  \label{fig:comtimeplot}
\end{figure}
which yields the estimated communalities in Figure~\ref{fig:comtimeplot}.

\begin{figure}[tp]
  \centering
  \includegraphics[width=0.49\textwidth]{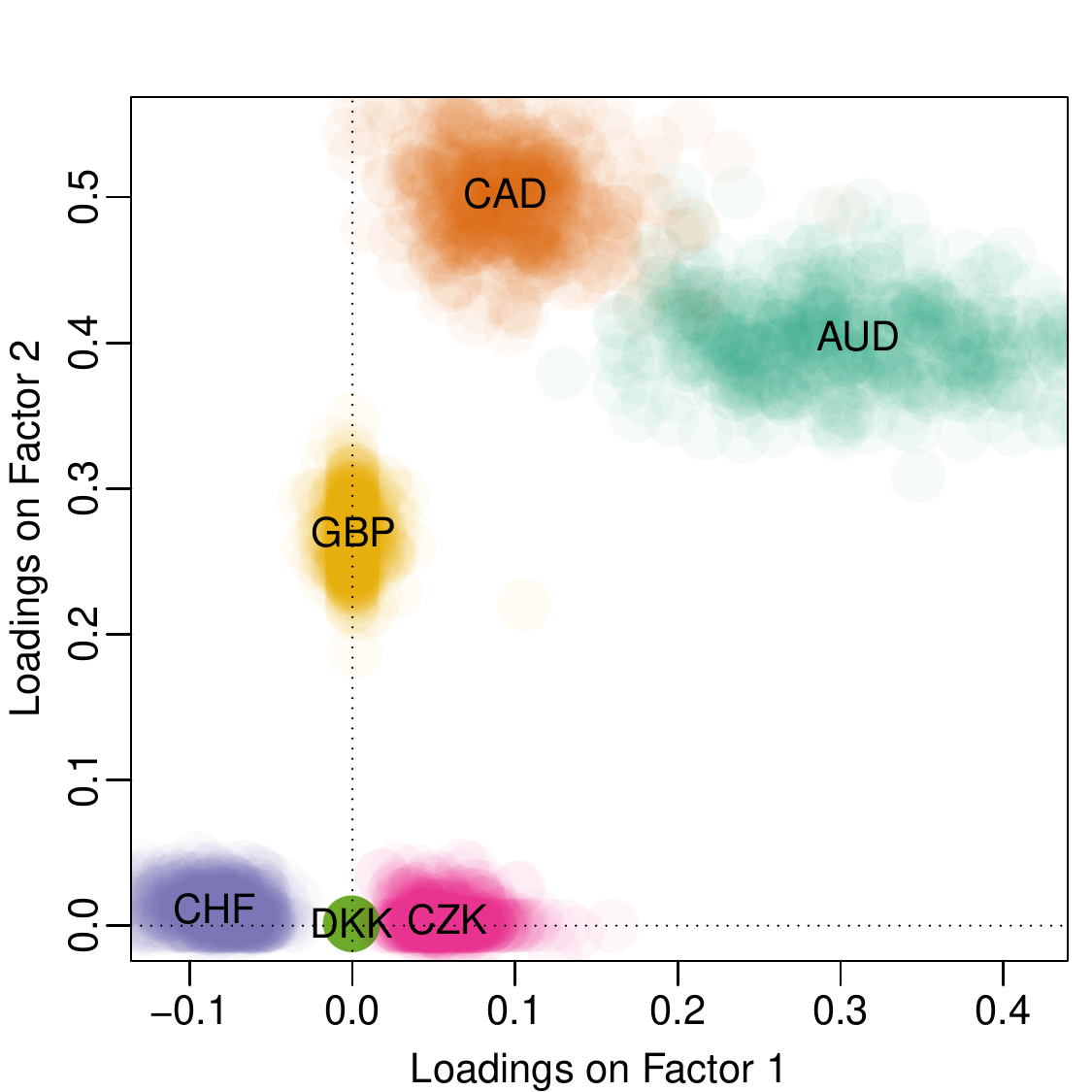}\hfill
  \includegraphics[width=0.49\textwidth]{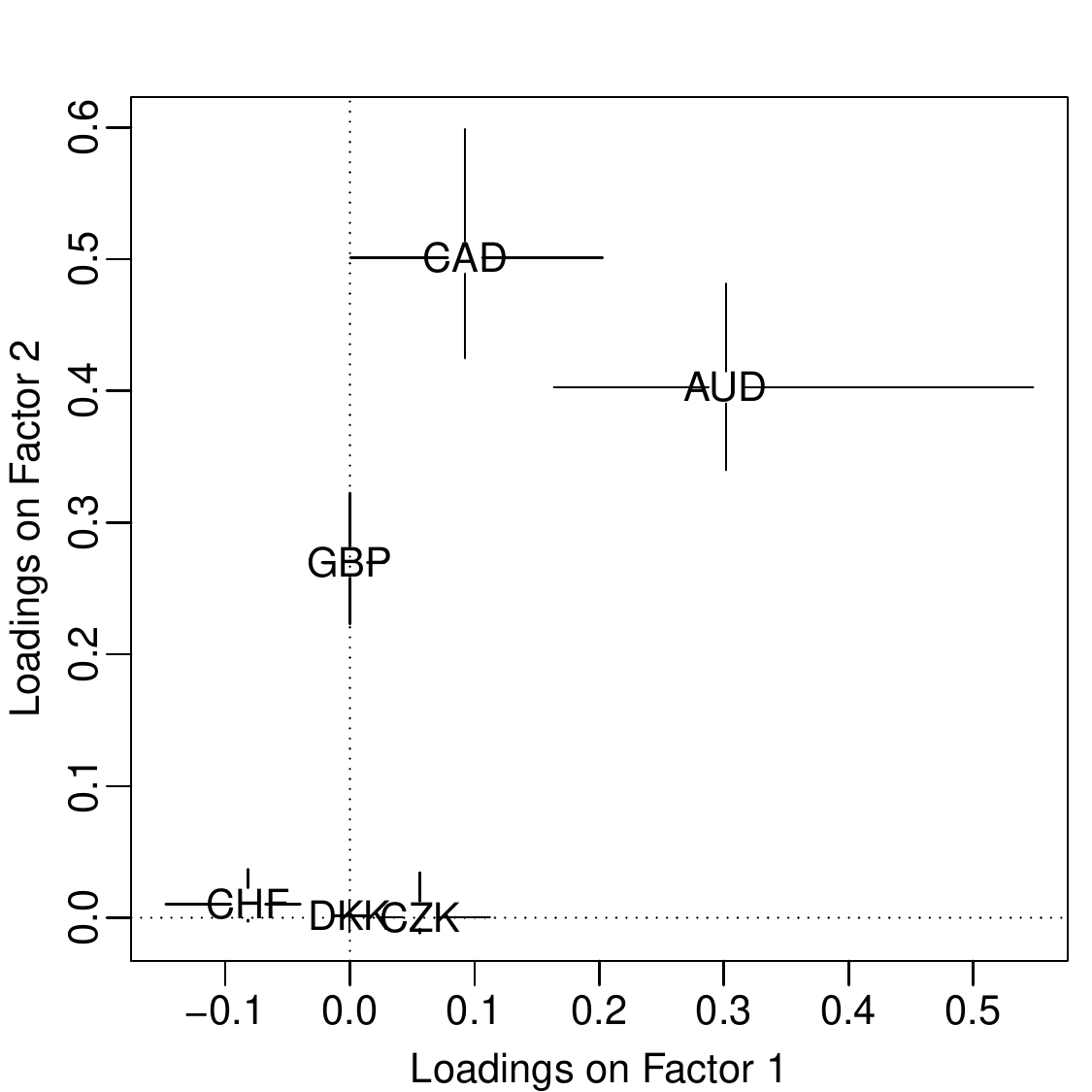}\\
  \includegraphics[width=\textwidth]{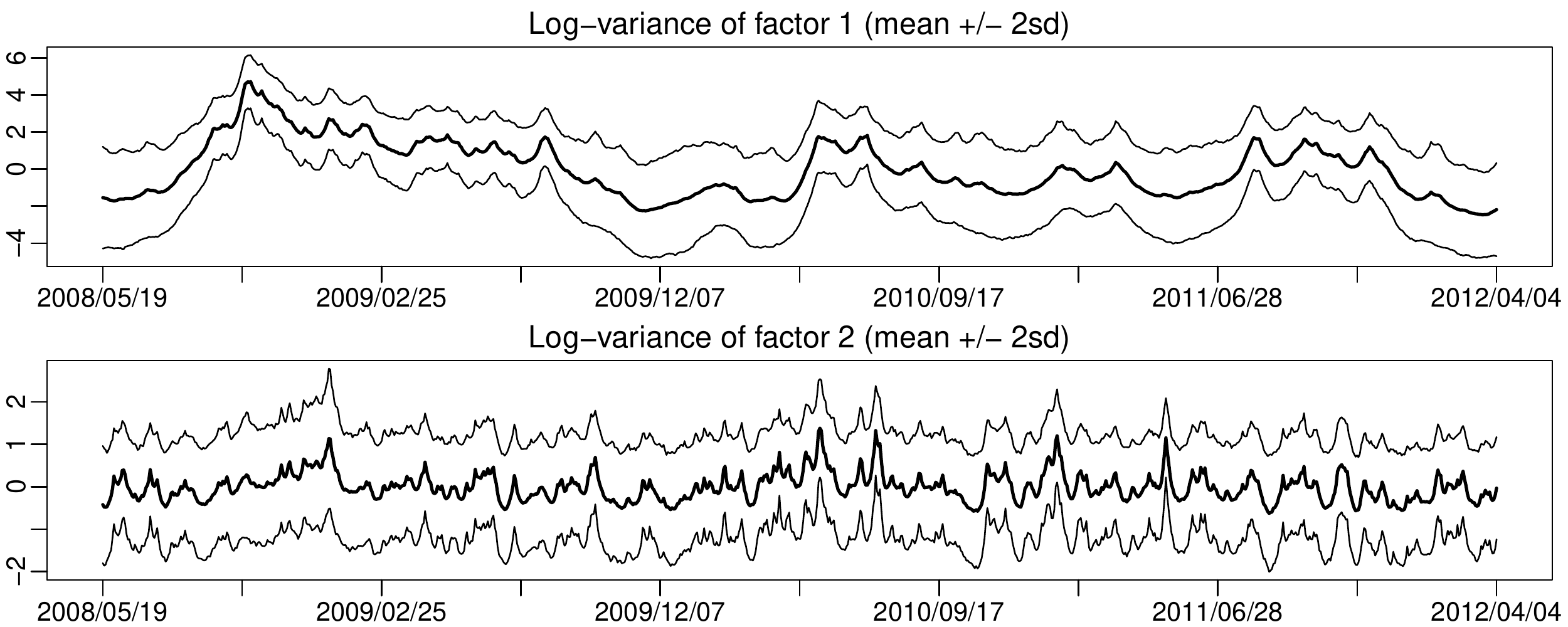}
  \caption{Factor loadings and factor variances.
    The top panels visualize the joint posterior distribution of the two factor loadings.
    While the colored clouds on the left consist of posterior draws and provide details, the posterior $0.01$/$0.99$ credible intervals on the right summarize the marginal distributions. 
    The bottom panels depict posterior means plus/minus two standard deviations of factor log variances.}
  \label{fig:loadplot2}
\end{figure}

\begin{figure}[tp]
  \centering
  \includegraphics[width=\textwidth]{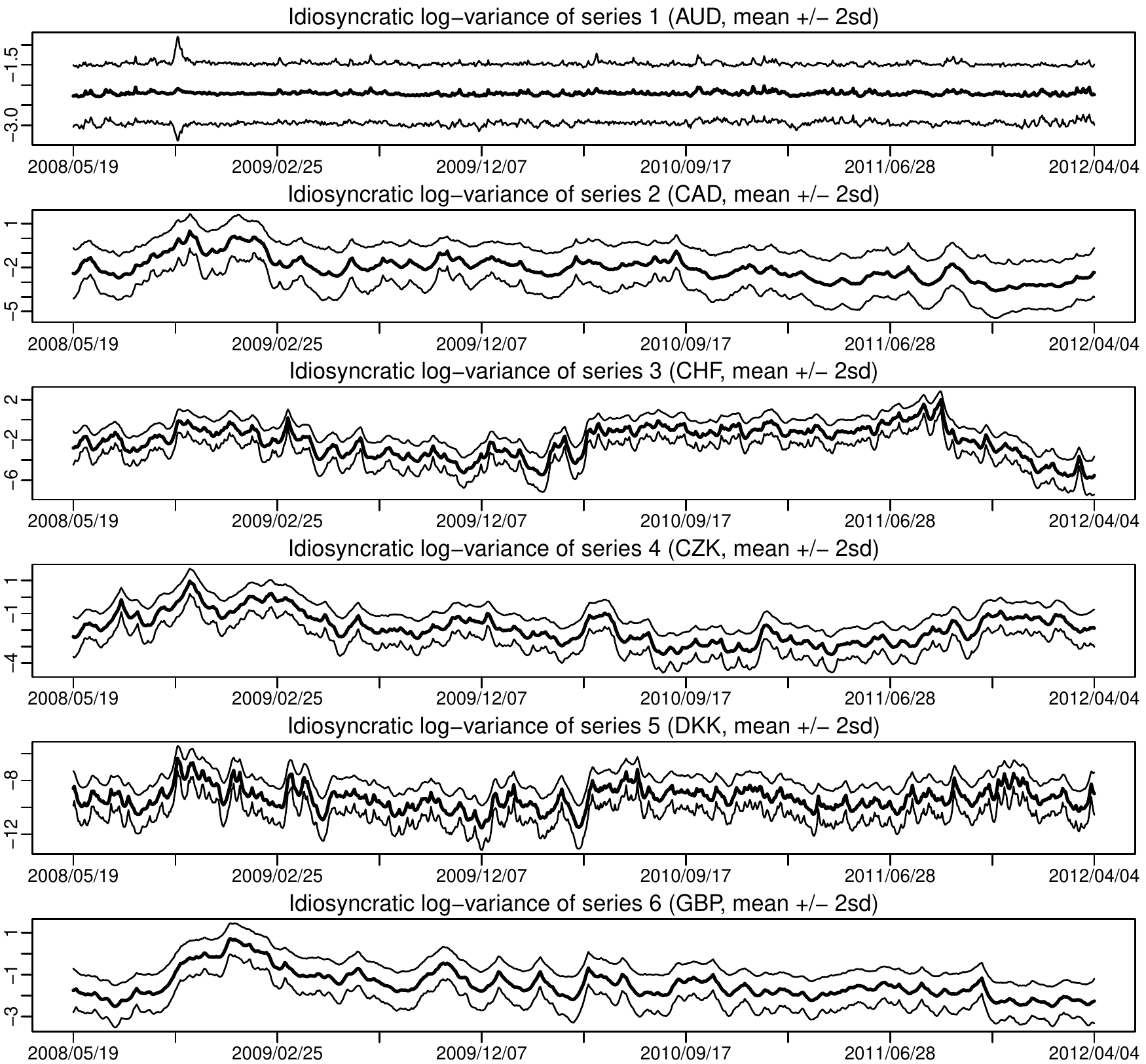}
  \caption{Idiosyncratic log variances: Posterior means plus/minus two standard deviations.
    Panels illustrate the marginal posterior distributions of the diagonal elements of $\Sigmaidi_t$ in Equation~\ref{eq:fsvcov}.}
  \label{fig:idivarplot}
\end{figure}

To gain an even deeper understanding of the estimated model, we now turn towards examining the latent factors and their variances themselves.
To visualize the loadings, the functions \fct{facloadpairplot}, \fct{facloadcredplot}, \fct{facloadpointplot}, \fct{facloadtraceplot}, and \fct{facloaddensplot} are available; the former two are exemplified in Figure~\ref{fig:loadplot2}.
Moreover, we can see the factor log variances produced through \code{logvartimeplot(res, show = "fac")}.
Similarly, \code{logvartimeplot(res, show = "idi")} produces plots of the idiosyncratic log variances which are displayed in Figure~\ref{fig:idivarplot}.
\begin{knitrout}
\definecolor{shadecolor}{rgb}{0.969, 0.969, 0.969}\color{fgcolor}\begin{kframe}
\begin{alltt}
\hlstd{R> }\hlkwd{facloadpairplot}\hlstd{(res)}
\hlstd{R> }\hlkwd{facloadcredplot}\hlstd{(res)}
\end{alltt}
\end{kframe}
\end{knitrout}
\begin{knitrout}
\definecolor{shadecolor}{rgb}{0.969, 0.969, 0.969}\color{fgcolor}\begin{kframe}
\begin{alltt}
\hlstd{R> }\hlkwd{logvartimeplot}\hlstd{(res,} \hlkwc{show} \hlstd{=} \hlstr{"fac"}\hlstd{)}
\end{alltt}
\end{kframe}
\end{knitrout}
\begin{knitrout}
\definecolor{shadecolor}{rgb}{0.969, 0.969, 0.969}\color{fgcolor}\begin{kframe}
\begin{alltt}
\hlstd{R> }\hlkwd{logvartimeplot}\hlstd{(res,} \hlkwc{show} \hlstd{=} \hlstr{"idi"}\hlstd{,} \hlkwc{maxrows} \hlstd{=} \hlnum{6}\hlstd{)}
\end{alltt}
\end{kframe}
\end{knitrout}

In addition to the above, there is the plotting function \fct{paratraceplot} which produces trace plots of all parameters associated with the log variances processes: mean, persistence, and volatility of log variances.

\begin{figure}[tp]
  \centering
  \includegraphics[width=\textwidth]{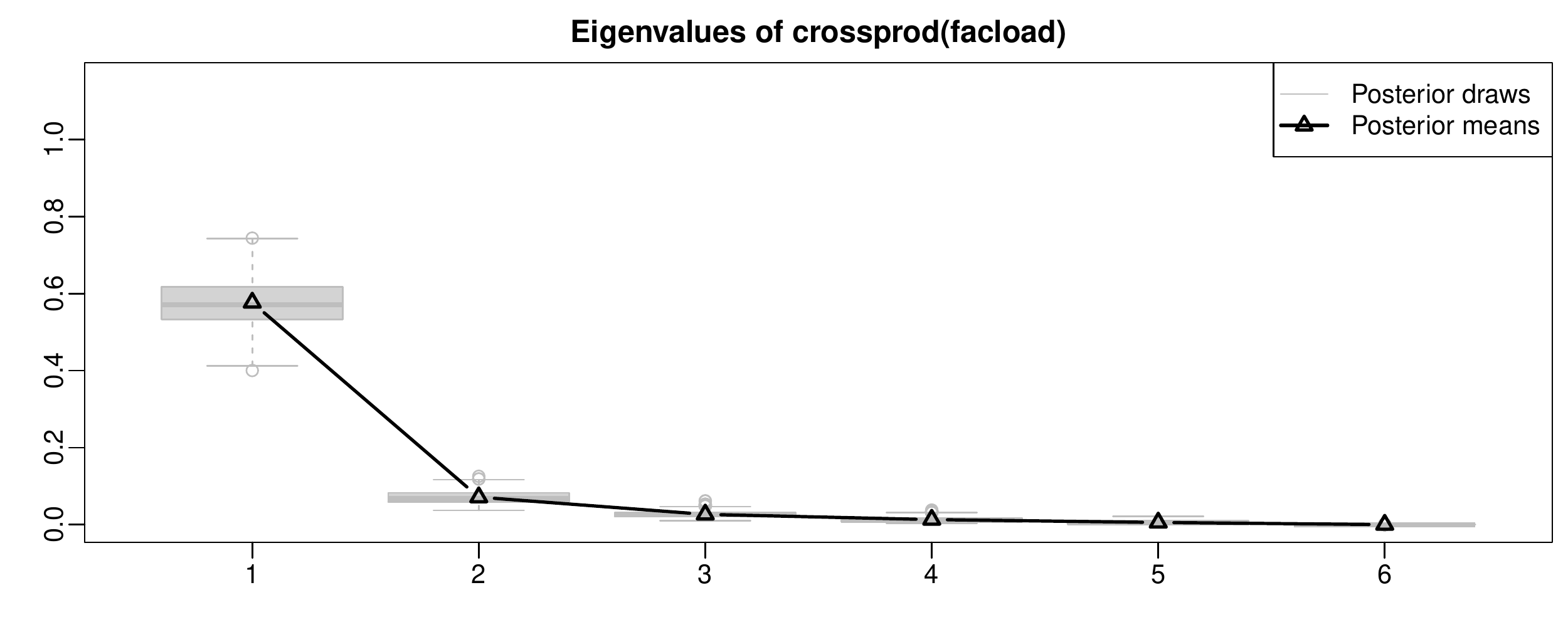}
  \caption{Eigenvalues of $\Loadings^\top\Loadings$ which can be used as a rough guide to selecting the number of factors.}
  \label{fig:evdiag}
\end{figure}

In order to provide some guidance when it comes to selecting the number of factors, \factorstochvol{} ships with the function \fct{evdiag}.
It computes and visualizes the eigenvalues of $\Loadings^\top\Loadings$ which can be used as a rough guide in analogy to a scree plot for static factor models.
To use it, one can fit a model with a relatively large number of factors and assess the importance of each of these (in descending order).
For example, the code below produces Figure~\ref{fig:evdiag}, hinting at a model with no more than four factors.

\begin{knitrout}
\definecolor{shadecolor}{rgb}{0.969, 0.969, 0.969}\color{fgcolor}\begin{kframe}
\begin{alltt}
\hlstd{R> }\hlkwd{set.seed}\hlstd{(}\hlnum{6}\hlstd{)}
\hlstd{R> }\hlstd{largemodel} \hlkwb{<-} \hlkwd{fsvsample}\hlstd{(y,} \hlkwc{factors} \hlstd{=} \hlnum{6}\hlstd{)}
\hlstd{R> }\hlkwd{evdiag}\hlstd{(largemodel)}
\end{alltt}
\end{kframe}
\end{knitrout}

\subsection{Predicting covariances, correlations, and future observations}

One of the main use cases of \factorstochvol{} might be to predict covariance and correlation matrices of time series. To this end, \fct{predcov} and \fct{predcor} yield draws from the posterior predictive distribution of these, defined in analogy to Equation~\ref{eq:predictive-density}. For instance, the code below can be used to obtain one-step-ahead posterior predictive means and standard deviations for the correlation matrix on April 5, 2012 (using data up to April 4 only).
\begin{knitrout}
\definecolor{shadecolor}{rgb}{0.969, 0.969, 0.969}\color{fgcolor}\begin{kframe}
\begin{alltt}
\hlstd{R> }\hlkwd{set.seed}\hlstd{(}\hlnum{4}\hlstd{)}
\hlstd{R> }\hlstd{predcor1} \hlkwb{<-} \hlkwd{predcor}\hlstd{(res)}
\hlstd{R> }\hlkwd{round}\hlstd{(}\hlkwd{apply}\hlstd{(predcor1[,,,}\hlnum{1}\hlstd{],} \hlnum{1}\hlopt{:}\hlnum{2}\hlstd{, mean),} \hlnum{2}\hlstd{)}
\end{alltt}
\begin{verbatim}
##      AUD  CAD   CHF   CZK  DKK  GBP
## AUD 1.00 0.62  0.00  0.03 0.10 0.47
## CAD 0.62 1.00  0.09  0.02 0.11 0.52
## CHF 0.00 0.09  1.00 -0.03 0.03 0.10
## CZK 0.03 0.02 -0.03  1.00 0.00 0.01
## DKK 0.10 0.11  0.03  0.00 1.00 0.09
## GBP 0.47 0.52  0.10  0.01 0.09 1.00
\end{verbatim}
\begin{alltt}
\hlstd{R> }\hlkwd{round}\hlstd{(}\hlkwd{apply}\hlstd{(predcor1[,,,}\hlnum{1}\hlstd{],} \hlnum{1}\hlopt{:}\hlnum{2}\hlstd{, sd),} \hlnum{2}\hlstd{)}
\end{alltt}
\begin{verbatim}
##      AUD  CAD  CHF  CZK  DKK  GBP
## AUD 0.00 0.16 0.17 0.04 0.07 0.16
## CAD 0.16 0.00 0.14 0.03 0.08 0.17
## CHF 0.17 0.14 0.00 0.04 0.04 0.11
## CZK 0.04 0.03 0.04 0.00 0.01 0.02
## DKK 0.07 0.08 0.04 0.01 0.00 0.07
## GBP 0.16 0.17 0.11 0.02 0.07 0.00
\end{verbatim}
\end{kframe}
\end{knitrout}

\begin{figure}[t]
  \centering
  \includegraphics[width=\textwidth, trim=20 20 20 40, clip]{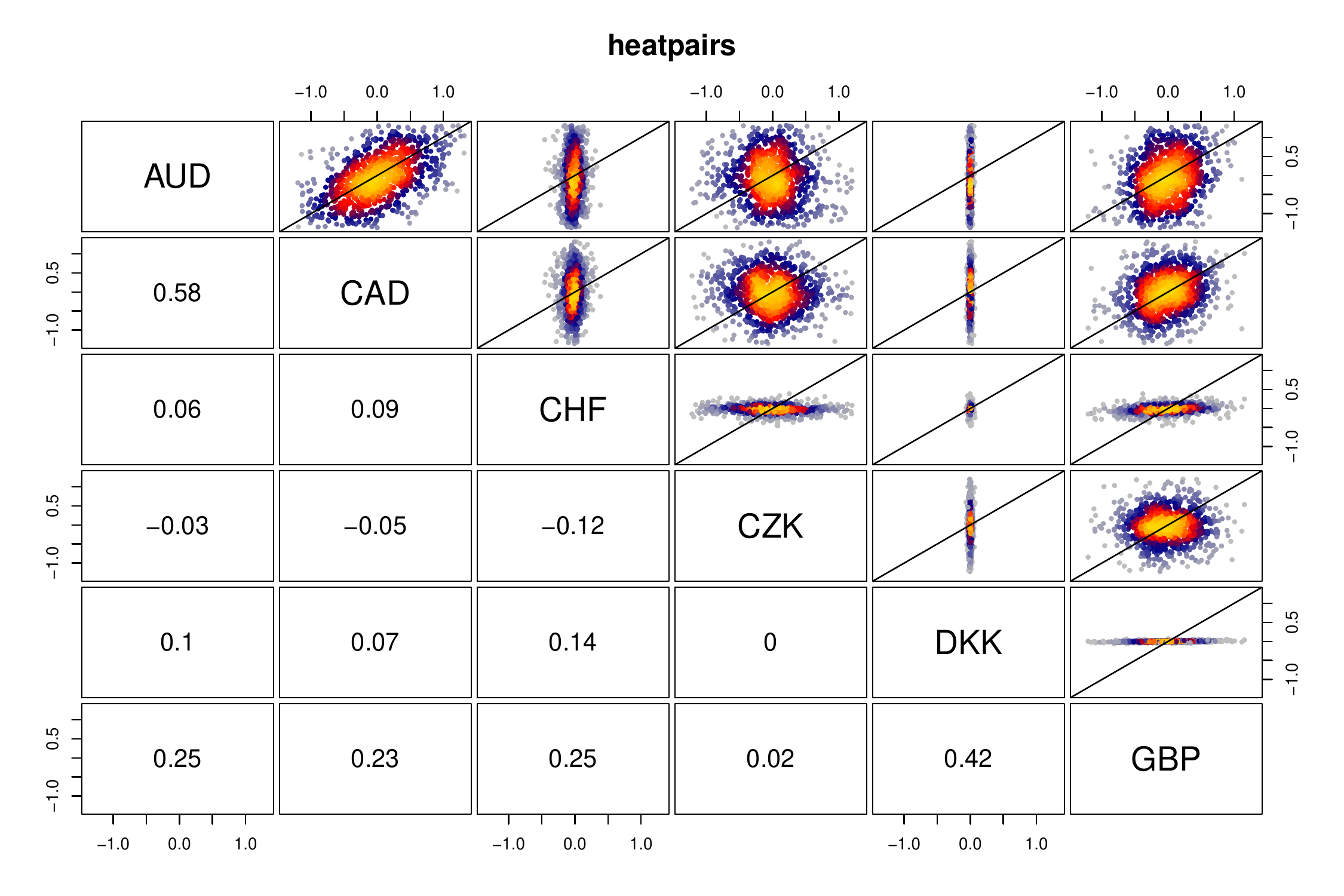}
  \caption{Predictive draws from the one-step-ahead posterior predictive distribution (above the diagonal) and empirical correlation coefficients (below the diagonal).}
  \label{fig:preddist}
\end{figure}

To obtain draws from the posterior predictive distribution of new data points, one can simply draw from the corresponding mixture of multivariate normals. In Figure~\ref{fig:preddist}, these draws are visualized via \fct{heatpairs} from \pkg{LSD} \citep{LSD}.
\begin{knitrout}
\definecolor{shadecolor}{rgb}{0.969, 0.969, 0.969}\color{fgcolor}\begin{kframe}
\begin{alltt}
\hlstd{R> }\hlkwd{set.seed}\hlstd{(}\hlnum{5}\hlstd{)}
\hlstd{R> }\hlstd{predcov_1} \hlkwb{<-} \hlkwd{predcov}\hlstd{(res)}
\hlstd{R> }\hlstd{preddraws} \hlkwb{<-} \hlkwd{t}\hlstd{(res}\hlopt{$}\hlstd{beta)}
\hlstd{R> }\hlkwa{for} \hlstd{(i} \hlkwa{in} \hlkwd{seq_len}\hlstd{(}\hlkwd{NROW}\hlstd{(preddraws)))}
\hlstd{+  }  \hlstd{preddraws[i,]} \hlkwb{<-} \hlstd{preddraws[i,]} \hlopt{+} \hlkwd{chol}\hlstd{(predcov_1[,,i,}\hlnum{1}\hlstd{])} \hlopt{%*%} \hlkwd{rnorm}\hlstd{(m)}
\hlstd{R> }\hlstd{plotlims} \hlkwb{<-} \hlkwd{quantile}\hlstd{(preddraws,} \hlkwd{c}\hlstd{(}\hlnum{0.01}\hlstd{,} \hlnum{0.99}\hlstd{))}
\hlstd{R> }\hlstd{LSD}\hlopt{::}\hlkwd{heatpairs}\hlstd{(preddraws,} \hlkwc{labels} \hlstd{=} \hlkwd{colnames}\hlstd{(y),} \hlkwc{cor.cex} \hlstd{=} \hlnum{1.5}\hlstd{,} \hlkwc{gap} \hlstd{=} \hlnum{0.3}\hlstd{,}
\hlstd{+  }  \hlkwc{xlim} \hlstd{= plotlims,} \hlkwc{ylim} \hlstd{= plotlims)}
\end{alltt}
\end{kframe}
\end{knitrout}

To conclude, we note that convenience functions such as \fct{predloglik} or \fct{predloglikWB} may aid in approximating predictive likelihoods \citep[cf., e.g.,][]{geweke2010comparing}. For instance, assuming that the actually observed value of $\bm y_{n+1} = \bm y_{n+2} = (0,0,0,0,0,0)^\top$, we can approximate the one and two step ahead log predictive scores through code along the following lines.
\begin{knitrout}
\definecolor{shadecolor}{rgb}{0.969, 0.969, 0.969}\color{fgcolor}\begin{kframe}
\begin{alltt}
\hlstd{R> }\hlkwd{set.seed}\hlstd{(}\hlnum{6}\hlstd{)}
\hlstd{R> }\hlkwd{predloglik}\hlstd{(res,} \hlkwd{matrix}\hlstd{(}\hlnum{0}\hlstd{,} \hlkwc{nrow} \hlstd{=} \hlnum{2}\hlstd{,} \hlkwc{ncol} \hlstd{= m),} \hlkwc{ahead} \hlstd{=} \hlnum{1}\hlopt{:}\hlnum{2}\hlstd{,} \hlkwc{each} \hlstd{=} \hlnum{10}\hlstd{)}
\end{alltt}
\begin{verbatim}
##        1        2 
## 5.093530 5.059362
\end{verbatim}
\end{kframe}
\end{knitrout}
For a more detailed analysis of the out-of-sample performance of \factorstochvol{} for exchange rate data, we refer to \cite{kastner2017efficient}; for a predictive exercise on stock market data, please see \cite{kastner2019sparse}.

\section{Summary and discussion} \label{sec:summary}

We extended the work of~\citet{kastner2016dealing} to other SV models, including the univariate heavy-tailed SV, the SV model with leverage, and the multivariate factor SV model.
We showcased the features that are the most important to end users in \proglang{R}: estimation through the sampler functions, visualization, summary, and prediction methods.
Due to its more involved nature, however, we did not include the description of the \proglang{C++} interface.
Two functions called \fct{update\_fast\_sv} and \fct{update\_general\_sv} are exported and programmers have the possibility to access the samplers in \stochvol{} directly from \proglang{C++} after linking to the compiled package.
For usage examples, see the implementations of \factorstochvol{} or \pkg{shrinkTVP} \citep{shrinktvp}.
Source code is available in the GitHub repositories \url{https://github.com/gregorkastner/stochvol} and \url{https://github.com/gregorkastner/factorstochvol}.

\section*{Acknowledgments}

Gregor Kastner acknowledges funding from the Austrian Science Fund (FWF) for the project ``High-dimensional statistical learning: New methods to advance economic and sustainability policy'' (ZK 35), jointly carried out by the University of Klagenfurt, WU Vienna University of Economics and Business, Paris Lodron University Salzburg, TU Wien, and the Austrian Institute of Economic Research (WIFO).

\bibliography{ref}

\begin{thebibliography}{67}
\newcommand{\enquote}[1]{``#1''}
\providecommand{\natexlab}[1]{#1}
\providecommand{\url}[1]{\texttt{#1}}
\providecommand{\urlprefix}{URL }
\expandafter\ifx\csname urlstyle\endcsname\relax
  \providecommand{\doi}[1]{doi:\discretionary{}{}{}#1}\else
  \providecommand{\doi}{doi:\discretionary{}{}{}\begingroup
  \urlstyle{rm}\Url}\fi
\providecommand{\eprint}[2][]{\url{#2}}

\bibitem[{Abanto-Valle \emph{et~al.}(2017)Abanto-Valle, Langrock, Chen, and
  Cardoso}]{abanto2017maximum}
Abanto-Valle CA, Langrock R, Chen MH, Cardoso MV (2017).
\newblock \enquote{Maximum likelihood estimation for stochastic volatility in
  mean models with heavy-tailed distributions.}
\newblock \emph{Applied Stochastic Models in Business and Industry},
  \textbf{33}(4), 394--408.
\newblock \doi{10.1002/asmb.2246}.

\bibitem[{Aguilar and West(2000)}]{aguilar2000bayesian}
Aguilar O, West M (2000).
\newblock \enquote{Bayesian Dynamic Factor Models and Portfolio Allocation.}
\newblock \emph{Journal of Business and Economic Statistics}, \textbf{18}(3),
  338--357.
\newblock \doi{10.1080/07350015.2000.10524875}.

\bibitem[{Bollerslev(1986)}]{bollerslev1986garch}
Bollerslev T (1986).
\newblock \enquote{Generalized Autoregressive Conditional Heteroskedasticity.}
\newblock \emph{Journal of Econometrics}, \textbf{31}(3), 307--327.
\newblock \doi{10.1016/0304-4076(86)90063-1}.

\bibitem[{Bos(2012)}]{bos2012relating}
Bos CS (2012).
\newblock \enquote{Relating Stochastic Volatility Estimation Methods.}
\newblock In L~Bauwens, C~Hafner, S~Laurent (eds.), \emph{Handbook of
  Volatility Models and Their Applications}, pp. 147--174. John Wiley \& Sons.
\newblock \doi{10.1002/9781118272039.ch6}.

\bibitem[{Brooks \emph{et~al.}(2011)Brooks, Gelman, Jones, and
  Meng}]{brooks2011handbook}
Brooks S, Gelman A, Jones G, Meng XL (eds.) (2011).
\newblock \emph{Handbook of {M}arkov Chain {M}onte {C}arlo}.
\newblock Handbooks of Modern Statistical Methods, 1st edition. Chapman and
  Hall, CRC Press.
\newblock ISBN 9781420079418.

\bibitem[{Carter and Kohn(1994)}]{carter1994gibbs}
Carter CK, Kohn R (1994).
\newblock \enquote{On {G}ibbs Sampling for State Space Models.}
\newblock \emph{Biometrika}, \textbf{81}(3), 541--553.
\newblock \doi{10.1093/biomet/81.3.541}.

\bibitem[{Chib \emph{et~al.}(2006)Chib, Nardari, and
  Shephard}]{chib2006analysis}
Chib S, Nardari F, Shephard N (2006).
\newblock \enquote{Analysis of High Dimensional Multivariate Stochastic
  Volatility Models.}
\newblock \emph{Journal of Econometrics}, \textbf{134}(2), 341--371.
\newblock \doi{10.1016/j.jeconom.2005.06.026}.

\bibitem[{Creal(2017)}]{creal2017class}
Creal DD (2017).
\newblock \enquote{A Class of Non-Gaussian State Space Models With Exact
  Likelihood Inference.}
\newblock \emph{Journal of Business \& Economic Statistics}, \textbf{35}(4),
  585--597.
\newblock \doi{10.1080/07350015.2015.1092977}.

\bibitem[{Delatola and Griffin(2011)}]{delatola2011bayesian}
Delatola EI, Griffin JE (2011).
\newblock \enquote{Bayesian Nonparametric Modeling of the Return Distribution
  with Stochastic Volatility.}
\newblock \emph{Bayesian Analysis}, \textbf{6}, 901--926.
\newblock \doi{10.1214/11-BA632}.

\bibitem[{Eddelbuettel and Fran\c{c}ois(2011)}]{eddelbuettel2011rcpp}
Eddelbuettel D, Fran\c{c}ois R (2011).
\newblock \enquote{\pkg{Rcpp}: Seamless \proglang{R} and \proglang{C++}
  Integration.}
\newblock \emph{Journal of Statistical Software}, \textbf{40}(8), 1--18.
\newblock \doi{10.18637/jss.v040.i08}.

\bibitem[{Eddelbuettel and Sanderson(2014)}]{eddelbuettel2014rcpparmadillo}
Eddelbuettel D, Sanderson C (2014).
\newblock \enquote{\pkg{RcppArmadillo}: Accelerating \proglang{R} with
  High-Performance \proglang{C++} Linear Algebra.}
\newblock \emph{Computational Statistics and Data Analysis}, \textbf{71},
  1054--1063.
\newblock \doi{10.1016/j.csda.2013.02.005}.

\bibitem[{Engle(1982)}]{engle1982arch}
Engle RF (1982).
\newblock \enquote{Autoregressive Conditional Heteroskedasticity with Estimates
  of the Variance of {U}nited {K}ingdom Inflation.}
\newblock \emph{Econometrica}, \textbf{50}(4), 987--1007.
\newblock \doi{10.2307/1912773}.

\bibitem[{Fr{\"u}hwirth-Schnatter(1994)}]{fruhwirth1994data}
Fr{\"u}hwirth-Schnatter S (1994).
\newblock \enquote{Data Augmentation and Dynamic Linear Models.}
\newblock \emph{Journal of Time Series Analysis}, \textbf{15}(2), 183--202.
\newblock \doi{10.1111/j.1467-9892.1994.tb00184.x}.

\bibitem[{Fr{\"u}hwirth-Schnatter and Lopes(2018)}]{fruhwirth2018sparse}
Fr{\"u}hwirth-Schnatter S, Lopes HF (2018).
\newblock \enquote{Sparse Bayesian Factor Analysis when the Number of Factors
  is Unknown.}
\newblock \emph{arXiv e-prints}, arXiv:1804.04231.
\newblock \urlprefix\url{https://arxiv.org/abs/1804.04231}.

\bibitem[{Fr{\"u}hwirth-Schnatter and Wagner(2010)}]{fruhwirth2010stochastic}
Fr{\"u}hwirth-Schnatter S, Wagner H (2010).
\newblock \enquote{Stochastic Model Specification Search for {G}aussian and
  Partial Non-{G}aussian State Space Models.}
\newblock \emph{Journal of Econometrics}, \textbf{154}(1), 85--100.
\newblock \doi{10.1016/j.jeconom.2009.07.003}.

\bibitem[{Geweke(1993)}]{geweke1993bayesian}
Geweke J (1993).
\newblock \enquote{Bayesian Treatment of the Independent {S}tudent-$t$ Linear
  Model.}
\newblock \emph{Journal of Applied Econometrics}, \textbf{8}(S1), S19--S40.
\newblock \doi{10.1002/jae.3950080504}.

\bibitem[{Geweke(2004)}]{geweke2004getting}
Geweke J (2004).
\newblock \enquote{Getting It Right: {J}oint Distribution Tests of Posterior
  Simulators.}
\newblock \emph{Journal of the American Statistical Association}, \textbf{99},
  799--804.

\bibitem[{Geweke and Amisano(2010)}]{geweke2010comparing}
Geweke J, Amisano G (2010).
\newblock \enquote{Comparing and Evaluating {B}ayesian Predictive Distributions
  of Asset Returns.}
\newblock \emph{International Journal of Forecasting}, \textbf{26}(2),
  216--230.
\newblock \doi{10.1016/j.ijforecast.2009.10.007}.

\bibitem[{Ghalanos(2020)}]{rugarch}
Ghalanos A (2020).
\newblock \emph{\pkg{rugarch}: {U}nivariate {GARCH} Models}.
\newblock \proglang{R} package version 1.4-4,
  \urlprefix\url{https://CRAN.R-project.org/package=rugarch}.

\bibitem[{Ghysels \emph{et~al.}(1996)Ghysels, Harvey, and
  Renault}]{ghysels1996stochastic}
Ghysels E, Harvey AC, Renault E (1996).
\newblock \enquote{Stochastic Volatility.}
\newblock In GS~Maddala, CR~Rao (eds.), \emph{Handbook of Statistics},
  volume~14, chapter~5, pp. 119--191. Elsevier.
\newblock \doi{10.1016/s0169-7161(96)14007-4}.

\bibitem[{Griffin and Brown(2010)}]{griffin2010inference}
Griffin JE, Brown PJ (2010).
\newblock \enquote{Inference with Normal-Gamma Prior Distributions in
  Regression Problems.}
\newblock \emph{Bayesian Analysis}, \textbf{5}(1), 171--188.
\newblock \doi{10.1214/10-BA507}.

\bibitem[{Han(2006)}]{han2006asset}
Han Y (2006).
\newblock \enquote{Asset Allocation with a High Dimensional Latent Factor
  Stochastic Volatility Model.}
\newblock \emph{Review of Financial Studies}, \textbf{19}(1), 237--271.
\newblock \doi{10.1093/rfs/hhj002}.

\bibitem[{Harvey \emph{et~al.}(1994)Harvey, Ruiz, and
  Shephard}]{harvey1994multivariate}
Harvey AC, Ruiz E, Shephard N (1994).
\newblock \enquote{Multivariate Stochastic Variance Models.}
\newblock \emph{The Review of Economic Studies}, \textbf{61}(2), 247--264.
\newblock \doi{10.2307/2297980}.

\bibitem[{Harvey and Shephard(1996)}]{harvey1996estimation}
Harvey AC, Shephard N (1996).
\newblock \enquote{Estimation of an Asymmetric Stochastic Volatility Model for
  Asset Returns.}
\newblock \emph{Journal of Business {\&} Economic Statistics}, \textbf{14}(4),
  429--434.
\newblock \doi{10.1080/07350015.1996.10524672}.

\bibitem[{Higham(1990)}]{higham1990analysis}
Higham NJ (1990).
\newblock \enquote{Analysis of the {C}holesky Decomposition of a Semi-Definite
  Matrix.}
\newblock \emph{Technical report}, Manchester Institute for Mathematical
  Sciences.
\newblock MIMS EPrint 2008.56,
  \urlprefix\url{http://eprints.maths.manchester.ac.uk/1193/}.

\bibitem[{Hosszejni and Kastner(2019)}]{hosszejni2019approaches}
Hosszejni D, Kastner G (2019).
\newblock \enquote{Approaches Toward the {B}ayesian Estimation of the
  Stochastic Volatility Model with Leverage.}
\newblock In R~Argiento, D~Durante, S~Wade (eds.), \emph{Bayesian Statistics:
  New Challenges and New Generations}, pp. 75--83. Springer-Verlag.
\newblock \doi{10.1007/978-3-030-30611-3\_8}.

\bibitem[{Hosszejni and Kastner(2021)}]{kastner2020stochvol}
Hosszejni D, Kastner G (2021).
\newblock \emph{\pkg{stochvol}: Efficient Bayesian Inference for Stochastic
  Volatility ({SV}) Models}.
\newblock \proglang{R} package version 3.0.4,
  \urlprefix\url{http://CRAN.R-project.org/package=stochvol}.

\bibitem[{Hosszejni and Kastner(forthcoming)}]{thisjss}
Hosszejni D, Kastner G (forthcoming).
\newblock \enquote{Modeling Univariate and Multivariate Stochastic Volatility
  in \proglang{R} with \stochvol{} and \factorstochvol.}
\newblock \emph{Journal of Statistical Software}.
\newblock \urlprefix\url{https://arxiv.org/abs/1906.12123}.

\bibitem[{Huber(2015)}]{huber2015perfect}
Huber ML (2015).
\newblock \emph{Perfect Simulation}.
\newblock Monographs on Statistics and Applied Probability, 1st edition.
  Chapman and Hall, CRC Press.
\newblock ISBN 9781482232448.

\bibitem[{{ISO/IEC}(2017)}]{cpplanguage}
{ISO/IEC} (2017).
\newblock \emph{International Standard {ISO/IEC} 14882:2017({E}) --
  {P}rogramming Language \proglang{C++}}.
\newblock International Organization for Standardization (ISO).
\newblock \urlprefix\url{https://isocpp.org/std/the-standard}.

\bibitem[{Jacquier \emph{et~al.}(1994)Jacquier, Polson, and
  Rossi}]{jacquier1994bayesian}
Jacquier E, Polson NG, Rossi PE (1994).
\newblock \enquote{Bayesian Analysis of Stochastic Volatility Models.}
\newblock \emph{Journal of Business \& Economic Statistics}, \textbf{20}(1),
  69--87.
\newblock \doi{10.1080/07350015.1994.10524553}.

\bibitem[{Jacquier \emph{et~al.}(2004)Jacquier, Polson, and
  Rossi}]{jacquier2004bayesian}
Jacquier E, Polson NG, Rossi PE (2004).
\newblock \enquote{Bayesian Analysis of Stochastic Volatility Models with
  Fat-Tails and Correlated Errors.}
\newblock \emph{Journal of Econometrics}, \textbf{122}(1), 185--212.
\newblock \doi{10.1016/j.jeconom.2003.09.001}.

\bibitem[{Jensen and Maheu(2010)}]{jensen2010bayesian}
Jensen MJ, Maheu JM (2010).
\newblock \enquote{Bayesian Semiparametric Stochastic Volatility Modeling.}
\newblock \emph{Journal of Econometrics}, \textbf{157}(2), 306--316.
\newblock \doi{10.1016/j.jeconom.2010.01.014}.

\bibitem[{Jensen and Maheu(2014)}]{jensen2014estimating}
Jensen MJ, Maheu JM (2014).
\newblock \enquote{Estimating a Semiparametric Asymmetric Stochastic Volatility
  Model with a {D}irichlet Process Mixture.}
\newblock \emph{Journal of Econometrics}, \textbf{178}(3), 523--538.
\newblock \doi{0.1016/j.jeconom.2013.08.018}.

\bibitem[{Kastner(2015)}]{kastner2015heavy}
Kastner G (2015).
\newblock \enquote{Heavy-Tailed Innovations in the \proglang{R} Package
  \pkg{stochvol}.}
\newblock \emph{Technical report}, WU Vienna University of Economics and
  Business.
\newblock \urlprefix\url{https://epub.wu.ac.at/4918}.

\bibitem[{Kastner(2016)}]{kastner2016dealing}
Kastner G (2016).
\newblock \enquote{Dealing with Stochastic Volatility in Time Series Using the
  \proglang{R} Package \pkg{stochvol}.}
\newblock \emph{Journal of Statistical Software}, \textbf{69}(5), 1--30.
\newblock \doi{10.18637/jss.v069.i05}.

\bibitem[{Kastner(2019)}]{kastner2019sparse}
Kastner G (2019).
\newblock \enquote{Sparse {B}ayesian Time-Varying Covariance Estimation in Many
  Dimensions.}
\newblock \emph{Journal of Econometrics}, \textbf{210}(1), 98--115.
\newblock \doi{10.1016/j.jeconom.2018.11.007}.

\bibitem[{Kastner and Fr{\"{u}}hwirth-Schnatter(2014)}]{kastner2014ancillarity}
Kastner G, Fr{\"{u}}hwirth-Schnatter S (2014).
\newblock \enquote{Ancillarity-Sufficiency Interweaving Strategy ({ASIS}) for
  Boosting {MCMC} Estimation of Stochastic Volatility Models.}
\newblock \emph{Computational Statistics and Data Analysis}, \textbf{76},
  408--423.
\newblock \doi{10.1016/j.csda.2013.01.002}.

\bibitem[{Kastner \emph{et~al.}(2017)Kastner, Fr{\"u}hwirth-Schnatter, and
  Lopes}]{kastner2017efficient}
Kastner G, Fr{\"u}hwirth-Schnatter S, Lopes HF (2017).
\newblock \enquote{Efficient {B}ayesian Inference for Multivariate Factor
  Stochastic Volatility Models.}
\newblock \emph{Journal of Computational and Graphical Statistics},
  \textbf{26}(4), 905--917.
\newblock \doi{10.1080/10618600.2017.1322091}.

\bibitem[{Kastner and Hosszejni(2021)}]{factorstochvol}
Kastner G, Hosszejni D (2021).
\newblock \emph{\pkg{factorstochvol}: {B}ayesian Estimation of (Sparse) Latent
  Factor Stochastic Volatility Models}.
\newblock \proglang{R} package version 0.10.2,
  \urlprefix\url{https://cran.r-project.org/package=factorstochvol}.

\bibitem[{Kim \emph{et~al.}(1998)Kim, Shephard, and Chib}]{kim1998stochastic}
Kim S, Shephard N, Chib S (1998).
\newblock \enquote{Stochastic Volatility: Likelihood Inference and Comparison
  with {ARCH} Models.}
\newblock \emph{The Review of Economic Studies}, \textbf{65}(3), 361--393.
\newblock \doi{10.1111/1467-937x.00050}.

\bibitem[{Knaus \emph{et~al.}(2020)Knaus, Bitto-Nemling, Cadonna, and
  Frühwirth-Schnatter}]{shrinktvp}
Knaus P, Bitto-Nemling A, Cadonna A, Frühwirth-Schnatter S (2020).
\newblock \emph{\pkg{shrinkTVP}: Efficient Bayesian Inference for Time-Varying
  Parameter Models with Shrinkage}.
\newblock \proglang{R} package version 2.0.1,
  \urlprefix\url{https://CRAN.R-project.org/package=shrinkTVP}.

\bibitem[{Markowitz(1952)}]{markowitz1952portfolio}
Markowitz H (1952).
\newblock \enquote{Portfolio Selection.}
\newblock \emph{The Journal of Finance}, \textbf{7}(1), 77--91.
\newblock \doi{10.1111/j.1540-6261.1952.tb01525.x}.

\bibitem[{McCausland \emph{et~al.}(2011)McCausland, Miller, and
  Pelletier}]{mccausland2011simulation}
McCausland WJ, Miller S, Pelletier D (2011).
\newblock \enquote{Simulation Smoothing for State-Space Models: A Computational
  Efficiency Analysis.}
\newblock \emph{Computational Statistics and Data Analysis}, \textbf{55}(1),
  199--212.
\newblock \doi{10.1016/j.csda.2010.07.009}.

\bibitem[{McElreath(2015)}]{mcelreath2015statistical}
McElreath R (2015).
\newblock \emph{Statistical Rethinking: A {B}ayesian Course with Examples in
  \proglang{R} and \proglang{Stan}}.
\newblock Texts in Statistical Science, 1st edition. Chapman and Hall, CRC
  Press.
\newblock ISBN 9781482253443.

\bibitem[{Nakajima(2012)}]{nakajima2012bayesian}
Nakajima J (2012).
\newblock \enquote{Bayesian Analysis of Generalized Autoregressive Conditional
  Heteroskedasticity and Stochastic Volatility: Modeling Leverage, Jumps and
  Heavy-Tails for Financial Time Series.}
\newblock \emph{Japanese Economic Review}, \textbf{63}(1), 81--103.
\newblock \doi{10.1111/j.1468-5876.2011.00537.x}.

\bibitem[{Nakajima and Omori(2009)}]{nakajima2009leverage}
Nakajima J, Omori Y (2009).
\newblock \enquote{Leverage, Heavy-Tails and Correlated Jumps in Stochastic
  Volatility Models.}
\newblock \emph{Computational Statistics and Data Analysis}, \textbf{53}(6),
  2335--2353.
\newblock \doi{10.1016/j.csda.2008.03.015}.

\bibitem[{Nakajima and Omori(2012)}]{nakajima2012stochastic}
Nakajima J, Omori Y (2012).
\newblock \enquote{Stochastic Volatility Model with Leverage and Asymmetrically
  Heavy-Tailed Error Using {GH} Skew {S}tudent's $t$~Distribution.}
\newblock \emph{Computational Statistics \& Data Analysis}, \textbf{56}(11),
  3690--3704.
\newblock \doi{10.1016/j.csda.2010.07.012}.

\bibitem[{Neuwirth(2014)}]{rcolorbrewer}
Neuwirth E (2014).
\newblock \emph{\pkg{R{C}olor{B}rewer}: {C}olor{B}rewer Palettes}.
\newblock \proglang{R} package version 1.1-2,
  \urlprefix\url{https://CRAN.R-project.org/package=RColorBrewer}.

\bibitem[{Omori \emph{et~al.}(2007)Omori, Chib, Shephard, and
  Nakajima}]{omori2007stochastic}
Omori Y, Chib S, Shephard N, Nakajima J (2007).
\newblock \enquote{Stochastic Volatility with Leverage: Fast and Efficient
  Likelihood Inference.}
\newblock \emph{Journal of Econometrics}, \textbf{140}(2), 425--449.
\newblock \doi{10.1016/j.jeconom.2006.07.008}.

\bibitem[{Park and Casella(2008)}]{park2008bayesian}
Park T, Casella G (2008).
\newblock \enquote{The {B}ayesian {L}asso.}
\newblock \emph{Journal of the American Statistical Association},
  \textbf{103}(452), 681--686.
\newblock \doi{10.1198/016214508000000337}.

\bibitem[{Pitt and Shephard(1999)}]{pitt1999time}
Pitt MK, Shephard N (1999).
\newblock \enquote{Time-Varying Covariances: {A} Factor Stochastic Volatility
  Approach.}
\newblock In JM~Bernardo, JO~Berger, AP~Dawid, AFM Smith (eds.), \emph{Bayesian
  {S}tatistics 6 -- {P}roceedings of the {S}ixth {V}alencia {I}nternational
  {M}eeting}, pp. 547--570. Oxford University Press.

\bibitem[{Plummer \emph{et~al.}(2006)Plummer, Best, Cowles, and Vines}]{coda}
Plummer M, Best N, Cowles K, Vines K (2006).
\newblock \enquote{{CODA}: Convergence Diagnosis and Output Analysis for
  {MCMC}.}
\newblock \emph{\proglang{R} News}, \textbf{6}(1), 7--11.
\newblock \urlprefix\url{https://www.r-project.org/doc/Rnews/Rnews_2006-1.pdf}.

\bibitem[{{\proglang{R} Core Team}(2020)}]{rlanguage}
{\proglang{R} Core Team} (2020).
\newblock \emph{\proglang{R}: A Language and Environment for Statistical
  Computing}.
\newblock \proglang{R} Foundation for Statistical Computing, Vienna, Austria.
\newblock \urlprefix\url{https://www.R-project.org/}.

\bibitem[{Roberts and Rosenthal(2007)}]{roberts2005coupling}
Roberts GO, Rosenthal J (2007).
\newblock \enquote{Coupling and Ergodicity of Adaptive {M}arkov Chain {M}onte
  {C}arlo Algorithms.}
\newblock \emph{Journal of Applied Probability}, \textbf{44}(2), 458--475.
\newblock \doi{10.1239/jap/1183667414}.

\bibitem[{Rue(2001)}]{rue2001fast}
Rue H (2001).
\newblock \enquote{Fast Sampling of {G}aussian {M}arkov Random Fields.}
\newblock \emph{Journal of the Royal Statistical Society B}, \textbf{63}(2),
  325--338.
\newblock \doi{10.1111/1467-9868.00288}.

\bibitem[{Sanderson and Curtin(2016)}]{sanderson2016armadillo}
Sanderson C, Curtin R (2016).
\newblock \enquote{\pkg{Armadillo}: A Template-Based \proglang{C++} Library for
  Linear Algebra.}
\newblock \emph{Journal of Open Source Software}, \textbf{1}, 26--18.
\newblock \doi{10.21105/joss.00026}.

\bibitem[{Schwalb \emph{et~al.}(2018)Schwalb, Tresch, Torkler, Duemcke, Demel,
  Ripley, and Venables}]{LSD}
Schwalb B, Tresch A, Torkler P, Duemcke S, Demel C, Ripley B, Venables B
  (2018).
\newblock \emph{\pkg{LSD}: Lots of Superior Depictions}.
\newblock \proglang{R} package version 4.0-0,
  \urlprefix\url{https://CRAN.R-project.org/package=LSD}.

\bibitem[{Sentana and Fiorentini(2001)}]{sentana2001identification}
Sentana E, Fiorentini G (2001).
\newblock \enquote{Identification, Estimation and Testing of Conditionally
  Heteroskedastic Factor Models.}
\newblock \emph{Journal of Econometrics}, \textbf{102}(2), 143--164.
\newblock \doi{10.1016/S0304-4076(01)00051-3}.

\bibitem[{Silva \emph{et~al.}(2006)Silva, Lopes, and Migon}]{silva2006extented}
Silva RS, Lopes HF, Migon HS (2006).
\newblock \enquote{The Extended Generalized Inverse {G}aussian Distribution for
  Log-Linear and Stochastic Volatility Models.}
\newblock \emph{Brazilian Journal of Probability and Statistics},
  \textbf{20}(1), 67--91.
\newblock \urlprefix\url{https://www.jstor.org/stable/43601074}.

\bibitem[{Sisson \emph{et~al.}(2018)Sisson, Fan, and
  Beaumont}]{sisson2018handbook}
Sisson SA, Fan Y, Beaumont MA (eds.) (2018).
\newblock \emph{Handbook of Approximate {B}ayesian Computation}.
\newblock Handbooks of Modern Statistical Methods, 1st edition. Chapman and
  Hall, CRC Press.
\newblock ISBN 9781439881507.

\bibitem[{Taylor(1982)}]{taylor1982sv}
Taylor SJ (1982).
\newblock \enquote{Financial Returns Modeled by the Product of Two Stochastic
  Processes: A Study of Daily Sugar Prices 1961--75.}
\newblock In OD~Anderson (ed.), \emph{Time Series Analysis, Theory and
  Practice}, pp. 203--226. North-Holland, Amsterdam.

\bibitem[{Wahl(2020)}]{stochvolTMB}
Wahl JC (2020).
\newblock \emph{\pkg{stochvolTMB}: Likelihood Estimation of Stochastic
  Volatility Models}.
\newblock \proglang{R} package version 0.1.2,
  \urlprefix\url{https://CRAN.R-project.org/package=stochvolTMB}.

\bibitem[{Wickham \emph{et~al.}(2020)Wickham, Hester, and Chang}]{devtools}
Wickham H, Hester J, Chang W (2020).
\newblock \emph{\pkg{devtools}: Tools to Make Developing \proglang{R} Packages
  Easier}.
\newblock \proglang{R} package version 2.3.2,
  \urlprefix\url{https://CRAN.R-project.org/package=devtools}.

\bibitem[{Yu and Meng(2011)}]{yu2011to}
Yu Y, Meng XL (2011).
\newblock \enquote{To Center or not to Center: That is not the Question---{A}n
  Ancillarity-Sufficiency Interweaving Strategy ({ASIS}) for Boosting {MCMC}
  Efficiency.}
\newblock \emph{Journal of Computational and Graphical Statistics},
  \textbf{20}(3), 531--570.
\newblock \doi{10.1198/jcgs.2011.203main}.

\bibitem[{Zeileis and Grothendieck(2005)}]{zoo}
Zeileis A, Grothendieck G (2005).
\newblock \enquote{\pkg{zoo}: \proglang{S3} Infrastructure for Regular and
  Irregular Time Series.}
\newblock \emph{Journal of Statistical Software}, \textbf{14}(6), 1--27.
\newblock \doi{10.18637/jss.v014.i06}.

\bibitem[{Zhou \emph{et~al.}(2014)Zhou, Nakajima, and West}]{zhou2014bayesian}
Zhou X, Nakajima J, West M (2014).
\newblock \enquote{Bayesian Forecasting and Portfolio Decisions Using Dynamic
  Dependent Sparse Factor Models.}
\newblock \emph{International Journal of Forecasting}, \textbf{30}(4),
  963--980.
\newblock \doi{10.1016/j.ijforecast.2014.03.017}.

\end{thebibliography}

\end{document}